\documentclass[11pt]{amsart}

\usepackage[nocompress]{cite} %This right here is what puts the in-text citation numbers in order

\usepackage{subfigure}
\usepackage{caption}
\usepackage{amsaddr}
\usepackage{color}
\usepackage{enumerate}
\usepackage{bbm}
\usepackage[mathscr]{eucal}
\usepackage{verbatim}

\usepackage{hyperref}

\usepackage{graphicx}

\usepackage{mathtools}

\usepackage[margin=1in]{geometry}

\theoremstyle{remark}

\newcommand{\AAA}{\mathcal{A}}

\numberwithin{equation}{section}
\newcommand{\SW}{{\rm SW}}
\newcommand{\occ}{{\rm occ}}

\DeclareMathOperator*{\argmin}{arg\,min}

\begin{document}

\title[MetaPalette]{MetaPalette: A $k$-mer painting approach for metagenomic taxonomic profiling and quantification of novel strain variation.}

\author{David Koslicki$^1{}^*$, Daniel Falush$^2$}
\address{$^1$Mathematics Department, Oregon State University, Corvallis, OR 97330}
\address{$^2$ Institute of Life Sciences, University of Swansea, Singleton Park, Swansea, SA2 8PP UK}
\thanks{$^*$Corresponding Author: \url{david.koslicki@math.oregonstate.edu}}

\date{\today}

\begin{abstract}

Metagenomic profiling is challenging in part because of the highly uneven sampling of the tree of life by genome sequencing projects and the limitations imposed by performing phylogenetic inference at fixed taxonomic ranks. We present the algorithm {\sf MetaPalette} which uses long $k$-mer sizes ($k=30, 50$) to fit a $k$-mer ``palette'' of a given sample to the $k$-mer palette of reference organisms. By modeling the $k$-mer palettes of unknown organisms, the method also gives an indication of the presence, abundance, and evolutionary relatedness of novel organisms present in the sample. The method returns a traditional, fixed-rank taxonomic profile which is shown on independently simulated data to be one of the most accurate to date. Tree figures are also returned that quantify the relatedness of novel organisms to reference sequences and the accuracy of such figures is demonstrated on simulated spike-ins and a metagenomic soil sample.

The software implementing {\sf MetaPalette} is available at:
\begin{center}
\url{https://github.com/dkoslicki/MetaPalette}
\end{center} 
Pre-trained databases are included for Archaea, Bacteria, Eukaryota, and viruses.

\end{abstract}
\maketitle

\section{Introduction}
Metagenomics is a developing field used to characterize the organismal composition of microbial communities in environmental or clinical samples \cite{sharpton2014introduction}. A key step in most metagenomic analyses is to identify the organisms in the sample and their relative frequency.  A wide variety of different algorithms have been developed for this purpose. 

Most approaches, including the one described here, are based on relating sequenced reads to reference organism genome sequences. Conceptually, the aim of these approaches is to place the organisms in the sample on a ``tree of life" that has been defined in advance. In practice, the available reference organisms are extremely unevenly scattered through the true tree of life. Many medically important branches, such as enterobacteria, are relatively well sampled with many strains from the same species, while there are entire phyla of unculturable organisms that are unrepresented \cite{marcy2007dissecting,baker2010enigmatic,lan2012using,rinke2013insights}.

A further difficulty, both in theory and in practice, is that a fully resolved tree of life cannot be established, even from complete reference genomes. At the scale of individual species, homologous recombination scrambles variation so that a tree is not necessarily an appropriate representation of organismal relationships, while more distant phylogenetic relationships can be difficult to estimate due to the various technical challenges of reconstructing ancient evolutionary events \cite{puigbo2009search,wang2001limitations,philippe2011resolving,salichos2013inferring,roure2013impact}. 

Based on these practical considerations, an effective metagenomic method should both identify the closest organism or set of organisms in the reference set and also estimate the genetic difference between the closest reference(s) and the organism present in the sample. The method should work both if the closest neighbor is a distant member of the same phyla or if there are multiple strains within the species in question. Fine scale classification is important because the detailed knowledge we have of, for example, \textit{E. coli} shows that organisms from the same species can have entirely different ecology and phenotypic effects on their host \cite{kaper2004pathogenic}.

Given these difficulties, a number of different approaches are taken to characterize metagenomic samples.
A commonly used approach is first to place individual reads onto a tree constructed for a particular set of genes, and then attempt to sum the phylogenetic information across the reads \cite{haft2012high,nguyen2014tipp,sunagawa2013metagenomic,liu2010metaphyler}. Phylogenetic analysis of each read can be computationally challenging for large datasets and individual reads can often only be placed inaccurately. It is challenging to appropriately represent this uncertainty in later stages of the analysis. These approaches also break down if a tree is not a good representation of relationships amongst organisms, e.g. within species. Furthermore, while utilizing specific genes (so-called \textit{marker genes}) can increase computational efficiency, this approach throws away a considerable amount of information from sequences that do not align to the marker genes. As a result of these issues, these methods are typically accurate for genus level or higher classification but not for fine scale classification.  

Another approach identifies features that are characteristic of particular organisms, such as the frequency of $k$-mers \cite{wood2014kraken,ounit2015clark,koslicki2014wgsquikr}. These features are used either for taxonomic binning of individual reads or in order to compute the overall composition. Depending on the $k$-mer size utilized, these methods are either suitable only for higher level phylogenetic analysis (for small $k$-mers), or are highly dependent on the training database utilized (for larger $k$-mers). In either case, no existent method using this approach can accurately detect and classify organisms highly diverged from ones in the training database, and still struggle with quantifying strain-level variation. Using longer $k$-mers allows for higher specificity but using $k$-mers that are unique to specific taxa in the reference dataset (as in \cite{wood2014kraken,ounit2015clark}) ignores a great deal of information about evolutionary relatedness provided by other $k$-mers. It also makes the approach highly dependent on the specific composition of the reference dataset. We argue that utilizing all $k$-mers in a reference database and multiple $k$-mer sizes allows for the modeling of the $k$-mer signature of organisms absent from a given training database.

\begin{figure}[h!]
    \centering
    \includegraphics[trim={1.6cm 0 1.6cm 0},clip,scale=.6]{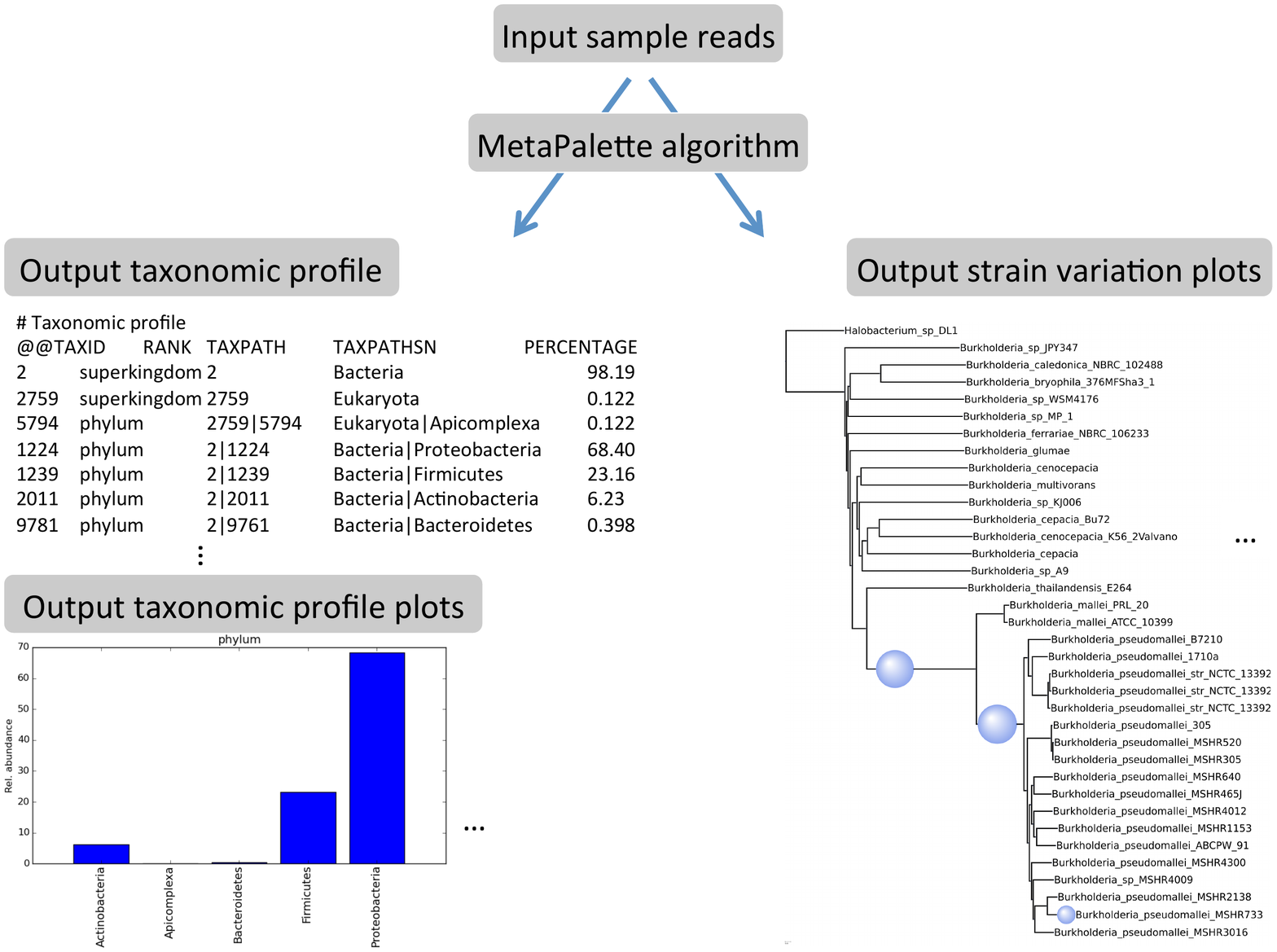}
    \caption{Illustration of the MetaPalette algorithm. Along with an output taxonomic profile and bar chart plots at all inferred taxonomic ranks, figures of strain-level variation for each inferred genus and/or species are also output.}
    \label{fig:workflow}
\end{figure}

In this manuscript, we present an approach based on defining a ``palette" for each reference organism. Specifically, we count the number of $k$-mers found in the sample DNA that are present in each reference organism. Our approach thus uses all $k$-mers of a particular length in the reference dataset, while discarding the specific information provided by matches of individual $k$-mers. This is similar in spirit to the so-called pseudoalignment approach of \cite{schaeffer2015pseudoalignment} except here we use $k$-mer counts of the entire sample, not of individual reads whose origins may be ambiguous. We model these palettes using a simple linear mixture model which includes both the reference organisms and ``hypothetical organisms" at different degrees of genetic relatedness to the reference organisms. The algorithm is called {\sf MetaPalette} and the outputs of the algorithm are demonstrated in Figure~\ref{fig:workflow}.

We first introduce the concept of a \textit{common $k$-mer matrix} and demonstrate how utilizing multiple $k$-mer sizes allows for accurate quantification of evolutionary relatedness. We then develop a mixture modeling procedure that utilizes this information to taxonomically profile a metagenomic sample and indicate the evolutionary relatedness of novel organisms. Evidence on simulated and real data is given that this approach can accurately capture strain-level variation, and we then benchmark this approach against other, commonly utilized metagenomic profiling techniques. 

\vspace{10ex}
\section{Methods}
\label{section:Methods}
\subsection{Common $k$-mer Training Matrix}
\label{section:CKM}
To quantify the similarity of two genomes, we count (with multiplicity) the fraction of each genome's $k$-mers that are in common with the other. Rigorous mathematical definitions of this and other quantities are contained in the Appendix section \ref{appendix:method}. This quantity, denoted ${\rm pckm}_k(\cdot,\cdot)$ for \textit{percent common $k$-mers}, is similar to the well-known Jaccard index \cite{jaccard1901etude} except that, among other differences, ${\rm pckm}_k(\cdot,\cdot)$ is not symmetric but does incorporate the counts of $k$-mers, not just their occurrence. 

\begin{figure}[h!]
    \centering
    \includegraphics[width=6in,height=4in]{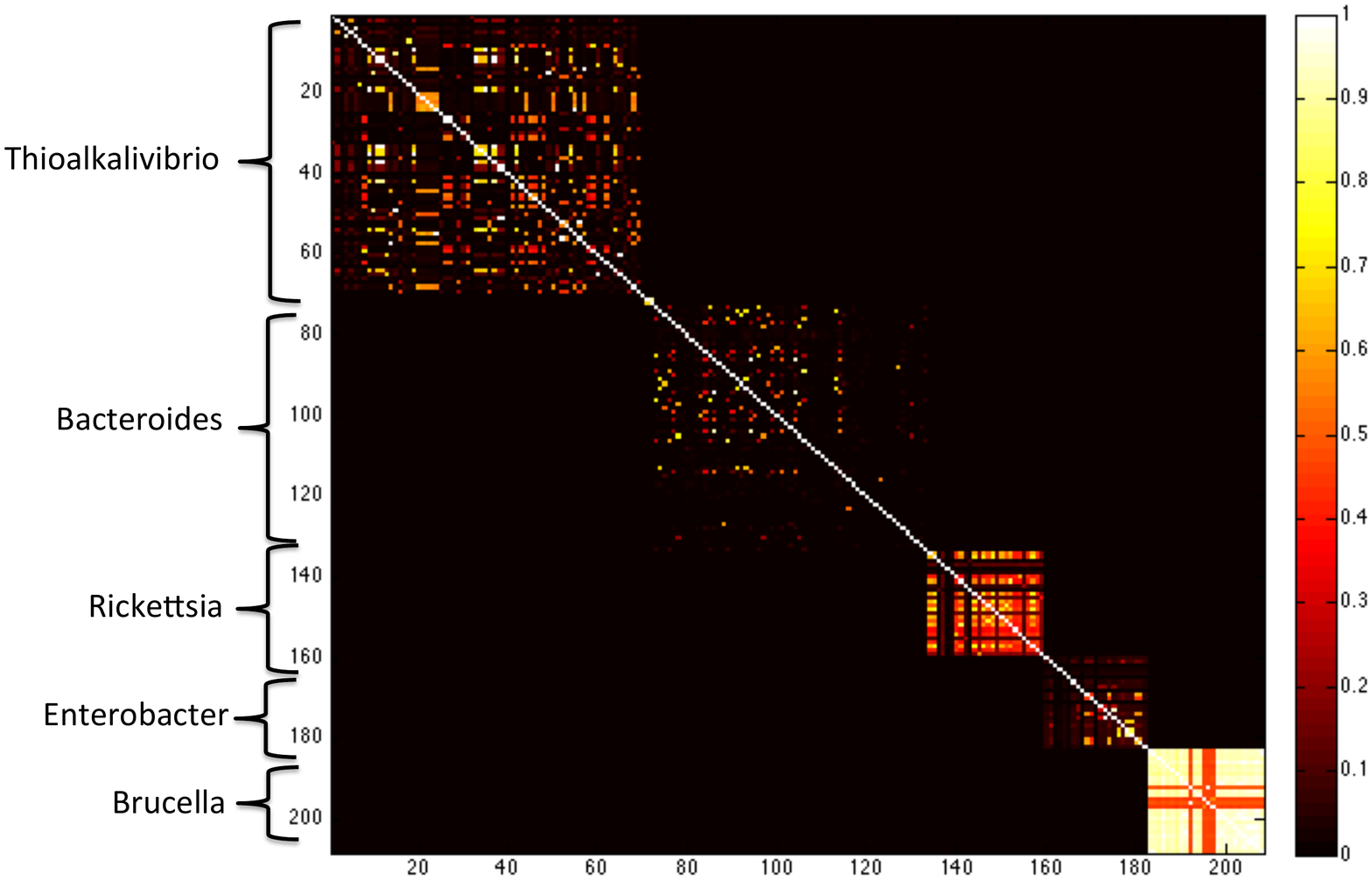} %HowSimilarAreTheGenera.m
    \caption{Heatmap of the common $k$-mer matrix $A^{(40)}$ for $k=40$ using a subset of the NCBI bacterial genome database. Delineations between genera can clearly be seen. In a given genera, differing similarity of species is also visible.}
    \label{fig:ckmtaxa}
\end{figure}

When given a set of genomes (i.e. a training database), a pair-wise similarity matrix can be formed: $A^{(k)}_{i,j} = {\rm pckm}_k(g_i,g_j)$ for $g_i$ and $g_j$ training genomes. The column vector ${\rm pckm}_k(\cdot,g_j)$ can be thought of as a palette, representing the particular $k$-mer profile of $g_i$ in relation to other genomes. We call each of these matrices a \textit{common $k$-mer matrix}. These matrices reflects the relatedness of the training genomes based on $k$-mer similarity. For larger $k$-mer sizes, one can clearly extract taxonomic information from these matrices: see Figure \ref{fig:ckmtaxa}.

\begin{figure}[h!]
    \centering
    \includegraphics[trim={1.6cm 2cm 3cm 4cm},clip,scale=.6]{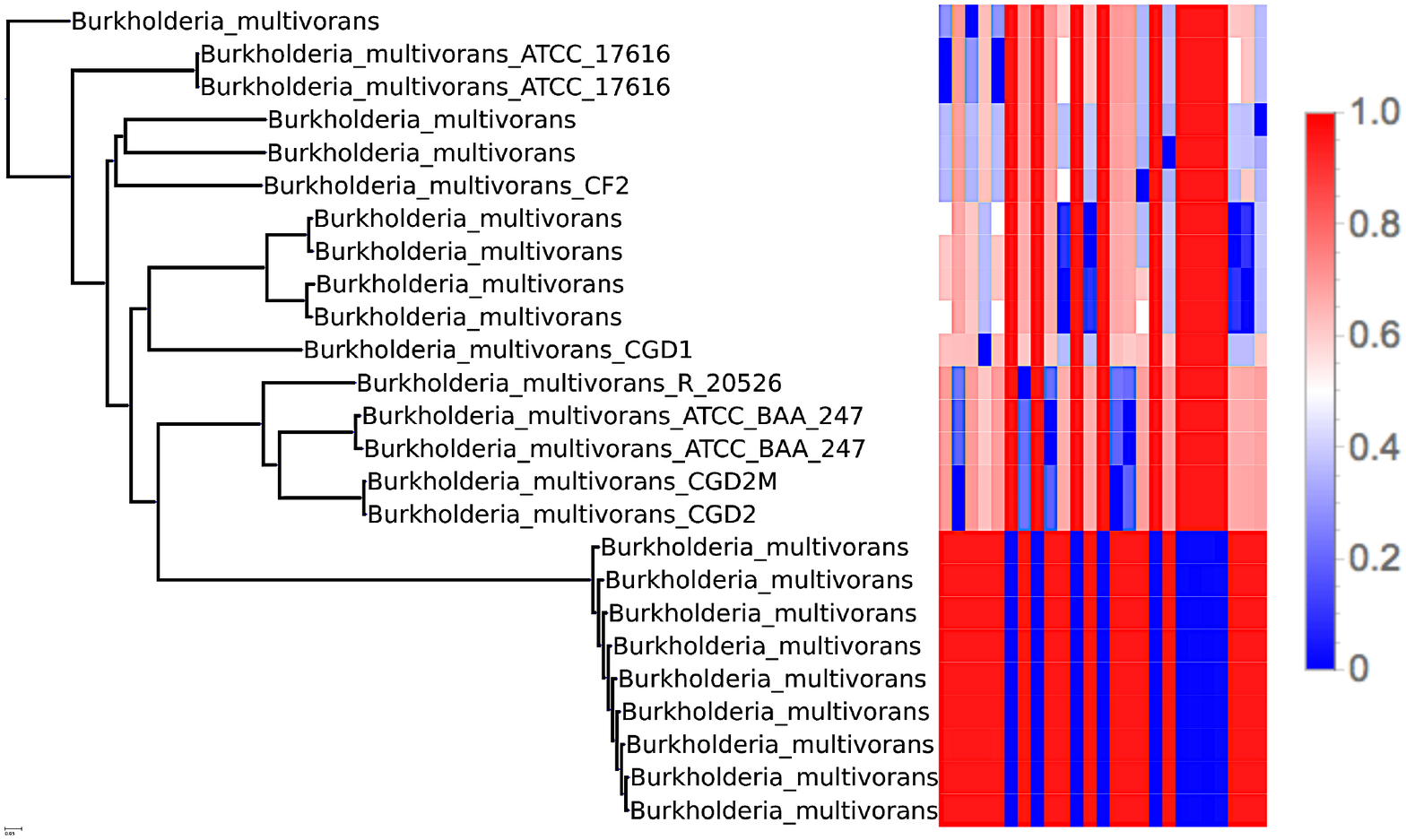} %pdf is with scale, png without
    \caption{Neighbor joining tree for the species \textit{Burkholderia multivorans} based on average of the common $30$-mer and $50$-mer matrices (shown in heat map to the right) depicting the ability of the common $k$-mer matrices to capture strain-level variation.}
    \label{fig:HeatTree}
\end{figure}

Beyond genera-level variation, strain-level variation can be captured through these common $k$-mer matrices. For example, using all the strains of the species \textit{Burkholderia multivorans} accessible via NCBI, we formed a neighbor joining tree using the average of the $30$-mer and $50$-mer common $k$-mer matrices. This tree, shown in figure \ref{fig:HeatTree}, demonstrates how the common $k$-mer matrices can capture variations amongst these strains.

The entries of $A^{(k)}$ can be calculated in a computationally efficient manner. We take the approach of forming bloom count filters (using Jellyfish \cite{marccais2011fast}) for each of the training genomes, and then counting the common $k$-mers using a simple C++ program based on heap data structures.

\subsection{Modeling Related Organisms}
\label{section:ModelingAbsentOrganisms}
\begin{figure*}[h!]
\centering
(a)
\begin{minipage}{3in}
\includegraphics[width=3in]{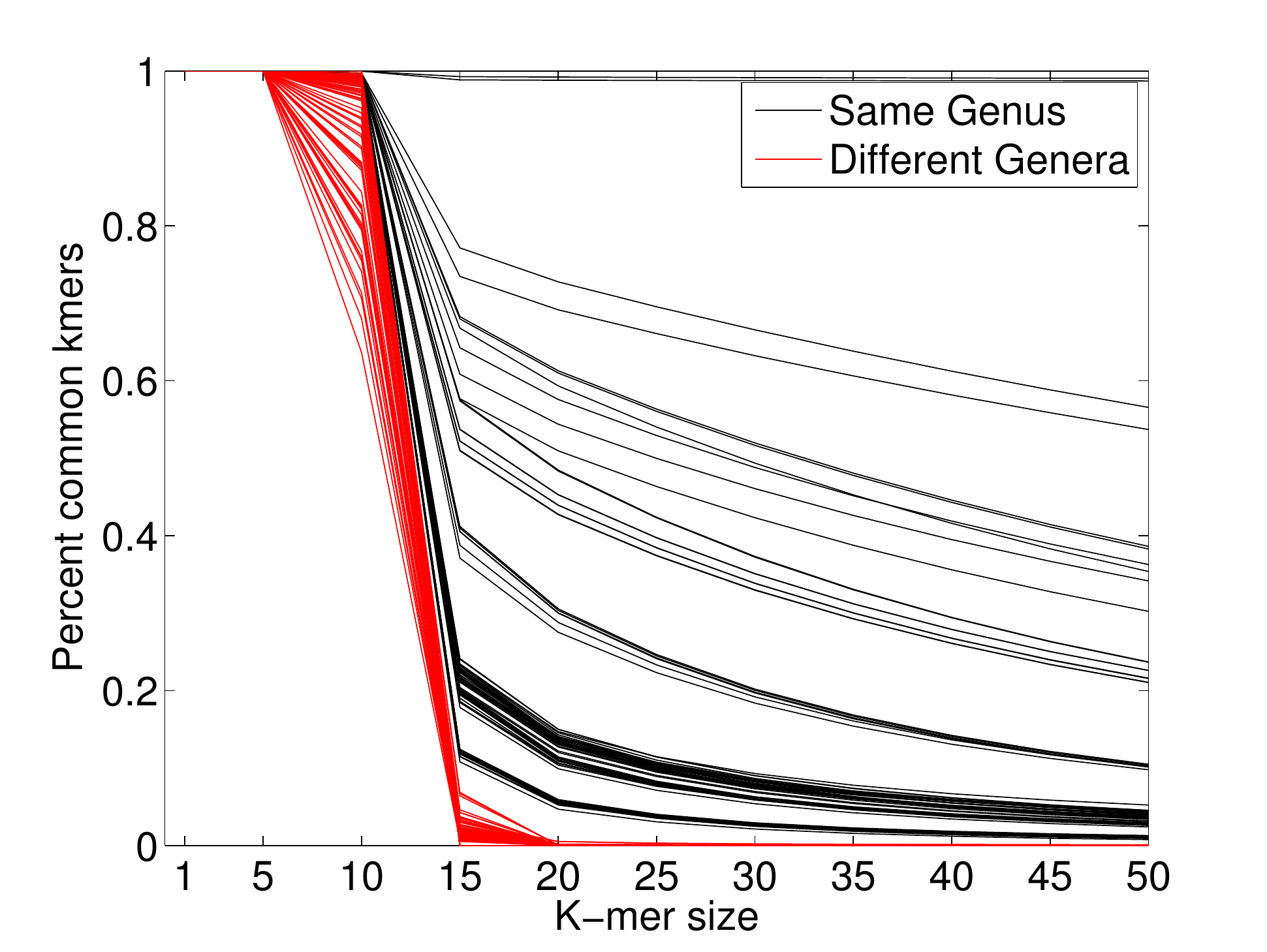}
% \caption{}
\end{minipage}
\begin{minipage}{3in}
(b)\begin{minipage}{3in}
\includegraphics[width=3in]{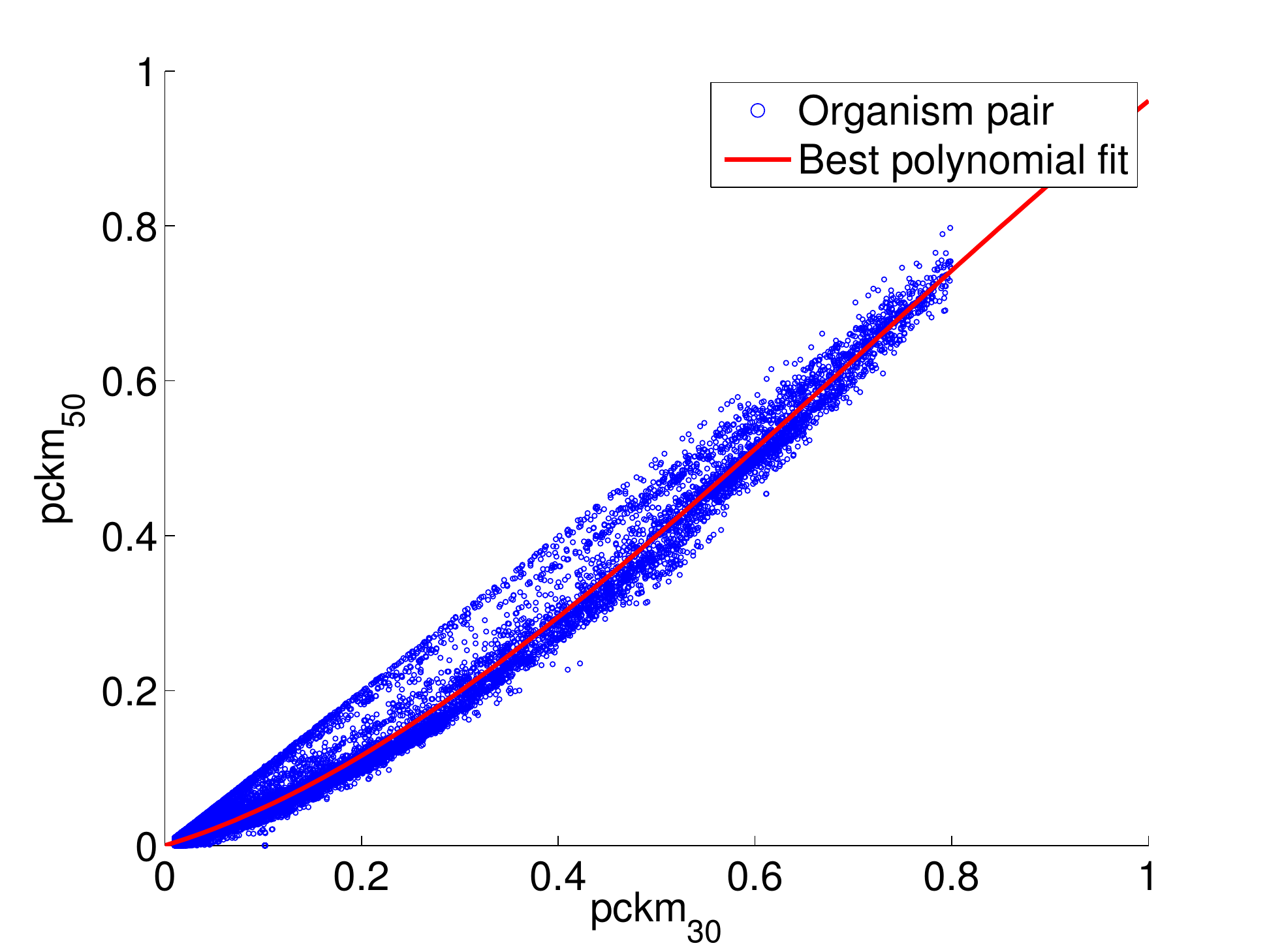}
% \caption{}
\end{minipage}
\end{minipage}
%\vspace{-5mm}
\caption{(a) Plot of $k$-mer similarity ${\rm pckm}_k(g_i,g_j)$ as a function of $k$ for 100 organism pairs of the same genus and 100 of different genera. (b) Scatter plot of the $6,914^2$ pairs of entries of the common $30$-mer and $50$-mer matricies shown with the best fit polynomial.}
\label{fig:ckmFunctionOfK}
\end{figure*}

To model the $k$-mer counts for organisms related at varying degrees from the training database, we take advantage of the differing behavior of ${\rm pckm}_k(\cdot,\cdot)$ as a function of $k$ for closely related organisms and distantly related organisms. In particular, the percent of common $k$-mers ${\rm pckm}_k(\cdot,\cdot)$ decays much slower as a function of $k$ for closely related organisms than for distantly related ones. This is consistent with the intuition that, for example, two organisms from different phyla will have a similar percent of shared $1$-mers, but very few common $50$-mers. Conversely, two closely related strains will have both a high percentage of shared $1$-mers \textit{and} a high percentage of shared $50$-mers. This is demonstrated in Figure \ref{fig:ckmFunctionOfK}(a). This property means that using more than one $k$-mer length should in principle allow us to distinguish between having an organism that is identical to a training organism at a low frequency and having an organism that is distantly related to all training organisms but present in the sample at a higher frequency. 

We focus on two particular $k$-mer sizes: $k=30$ and $k=50$ due to the predictability of ${\rm pckm}_k$ for these $k$-mer sizes. Indeed, using 6,914 whole bacterial genomes downloaded from a variety of publicly accessible repositories (via {\sf RepoPhlAn}: \url{https://bitbucket.org/nsegata/repophlan}), we observed that the percent of shared $30$-mers can be predicted from the percent of shared $50$-mers. See part (b) of Figure \ref{fig:ckmFunctionOfK}. A degree 3 polynomial was used (as it resulted in the lowest RMSE and ${\rm R}^2$ values, which did not improve for higher degree polynomials). Namely, we observed that for the polynomial $p(x)=-.5141x.^3+1.0932x.^2+0.3824x$, ${\rm pckm}_{50}(g_i,g_j) \approx p\left({\rm pckm}_{30}(g_i,g_j)\right)$.

For $k$-mer lengths substantially shorter than 30, the behavior of ${\rm pckm}_k$ is more variable, for example because of convergence of sequence composition between distantly related organisms. On the other hand, $k$-mers much larger than 50 are increasingly time consuming to compute and are likely to be more sensitive to sequencing error and other technical artifacts.

We can augment the matrices $A^{(k)}$ with columns that represent hypothetical organisms which are related by different degrees to the reference organism. For a given organism with genome $g_i$, if we wish to include a hypothetical organism $h$ that is 90\% similar to genome $g_i$ in its 30-mers, we can round down each entry of the column vector ${\rm pckm}_{30}(\cdot,g_i)$ to be no more than $0.90$. Call this vector ${\rm pckm}_{30}(\cdot,h)$. The entries below 90\% do not need to be changed since we assume that the hypothetical organism has the same patterns of $k$-mer sharing to more distantly related ``outgroup" taxa as to the reference organism. 

We model the $50$-mer similarity by setting ${\rm pckm}_{50}(\cdot,h)=p\left({\rm pckm}_{30}(\cdot,g_i)\right)$ for $p$ the previously defined polynomial. Adding these vectors to $A^{(30)}$ and $A^{(50)}$ effectively adds a hypothetical organism that has a common $k$-mer signature 90\% similar to genome $g_i$. We then repeat this procedure for all training genomes $g_i$ and  for similarities ranging from 90\%, 80\%,$\cdots$,10\% and append these columns to $A^{(30)}$ and $A^{(50)}$.

\subsection{Sample $k$-mer Signature}
\label{section:SampleKmerSignature}
Given a metagenomic sample, we form two vectors $y^{(30)}$ and  $y^{(50)}$ consisting of the total counts in the sample of the $30$-mers and $50$-mers shared with the training organisms. In the Appendix section \ref{appendix:MathFormulation}, we show that these vectors are linearly related to the organism abundances via the common $k$-mer matrices $A^{(30)}$ and $A^{(50)}$.

Note that in forming $y^{(k)}$, we count the $k$-mers in the entire sample, not of the individual reads. This allows for a very computationally efficient approach: as the training genomes typically have low error, their $k$-mers can be efficiently stored in de Bruijn graphs (formed using {\sf Bcalm} \cite{chikhi2014representation}). We can then query the bloom count filter formed from the sample in a highly parallel fashion.

\subsection{Sparsity Promoting Optimization Procedure}
\label{section:SparsityPromotingOptimizationProcedure}
After forming $y^{(k)}$, we note that some of the entries $y_i^{(k)}$ may be non-zero not due to the presence of organism $i$ in the sample, but due to the fact that there exists an organism $j$ that shares portions of its genome with organism $i$. Since $A_{i,j}^{(k)}$ represents the ``overlap'' of these two organisms, we can deconvolute this linear mixture relationship by solving the equation $A^{(k)} x = y^{(k)}$ for $x$ the vector of organism abundances. However, after having augmented $A^{(k)}$ with the hypothetical organisms, this system of equations is underdetermined (10 times more columns than rows). We can employ a sparsity promoting optimization procedure to infer the most parsimonious $x$ consistent with the equations $A^{(k)} x = y^{(k)}$ for $k=30,50$. This procedure, first introduced in \cite{koslicki2013quikr} and proven correct in \cite{FoucartKoslicki2014}, is detailed in the Appendix section \ref{section:OptimizationProcedure}.

\subsection{Inferring Taxonomy}
The abundances of the hypothetical organisms is then mapped back onto the taxonomy (for the output taxonomic profile) or the neighbor joining tree formed from the $A^{(k)}$ (for the output strain variation figures) utilizing a least common ancestor approach detailed in the Appendix section \ref{Appendix:InferringTaxonomy}.

\section{Quantification of Strain-level Variation}
\label{section:StrainVariation}

We demonstrate in two ways that the inclusion of the hypothetical organisms allows for the inference of strain-level variation. First, we spike novel organisms into a mock metagenomic community and show that {\sf MetaPalette} can accurately predict their presence. Second, we utilize a real metagenomic soil sample to give evidence for a novel strain that {\sf MetaPalette} predicts.

\subsection{HMP Mock Community}
\begin{figure}[ht!]
    \centering
	\includegraphics[trim={0cm 1cm 0cm 0cm},clip,scale=.14]{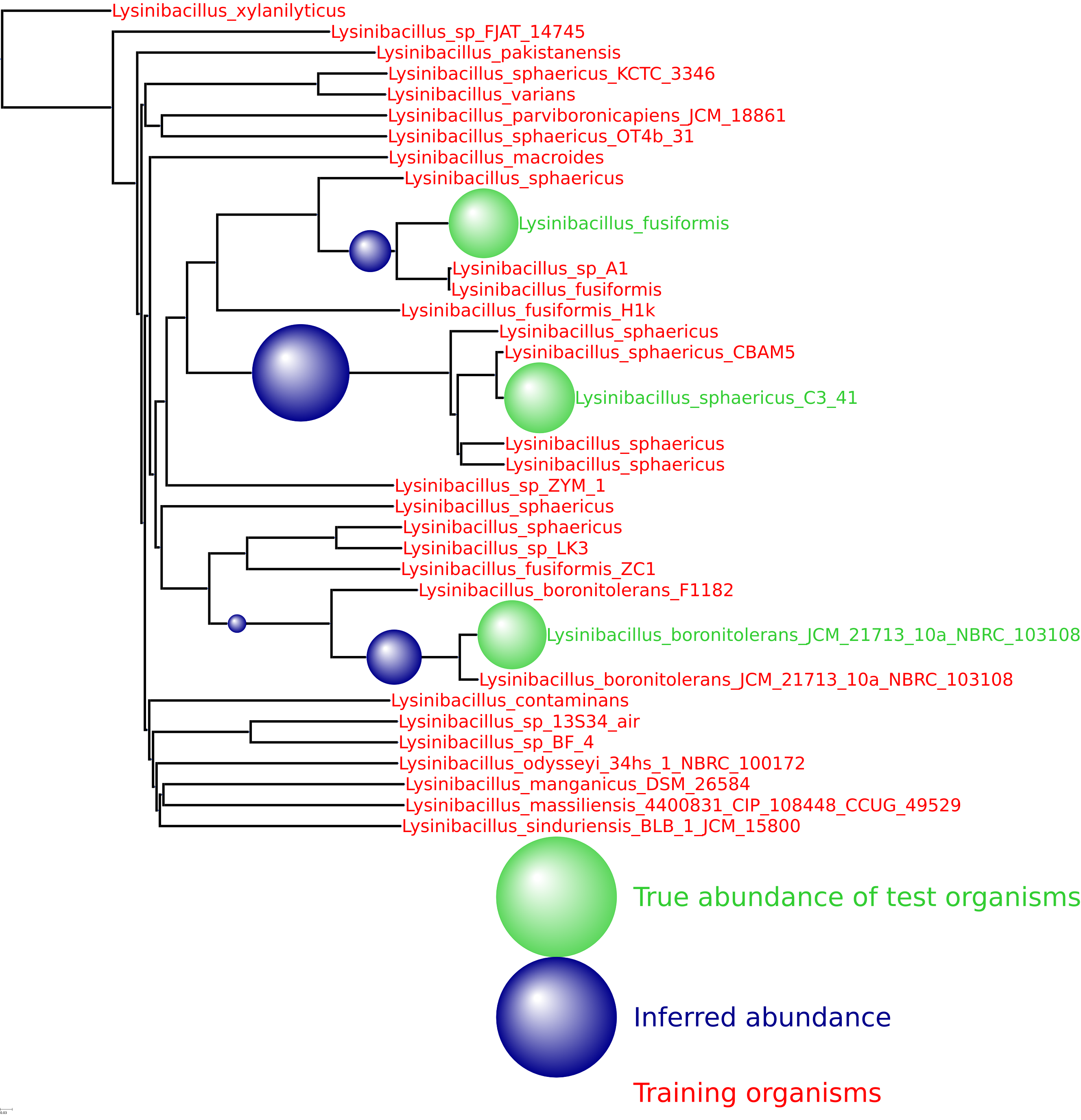}
    \caption{Result from training on 30 strains of \textit{L. sphaericus} and testing on three novel strains. A total of 50K reads from the novel strains were spiked into the HMP mock even community. The training organisms are denoted with red names and the testing organisms have green names. The true abundance of the sample is pictures with green spheres and the inferred abundance is pictured with blue spheres. }
    \label{fig:TwoAbsent}
\end{figure}

We first formed the common $k$-mer matrices $A^{(k)}$ using 31 strains of \textit{Lysinibacillus sphaericus}. We then used {\sf Grinder} \cite{Angly2012} to simulate a dataset consisting of two novel strains (not included in the training database). These reads where then spiked into the HMP mock even community (a $\sim$6.6M read metagenome consisting of 22 select organisms sampled using an Illumina GA-II sequencer; NCBI accession SRR172902). The output of {\sf MetaPalette} is shown in Figure \ref{fig:TwoAbsent} demonstrating the ability of the method to correctly infer the presence of organisms absent from the training data.

\begin{figure*}[h!]
\centering
\hspace{-5ex}
(a)
\begin{minipage}{3.5in}
\hspace{-1.5ex}
\includegraphics[width=3.5in]{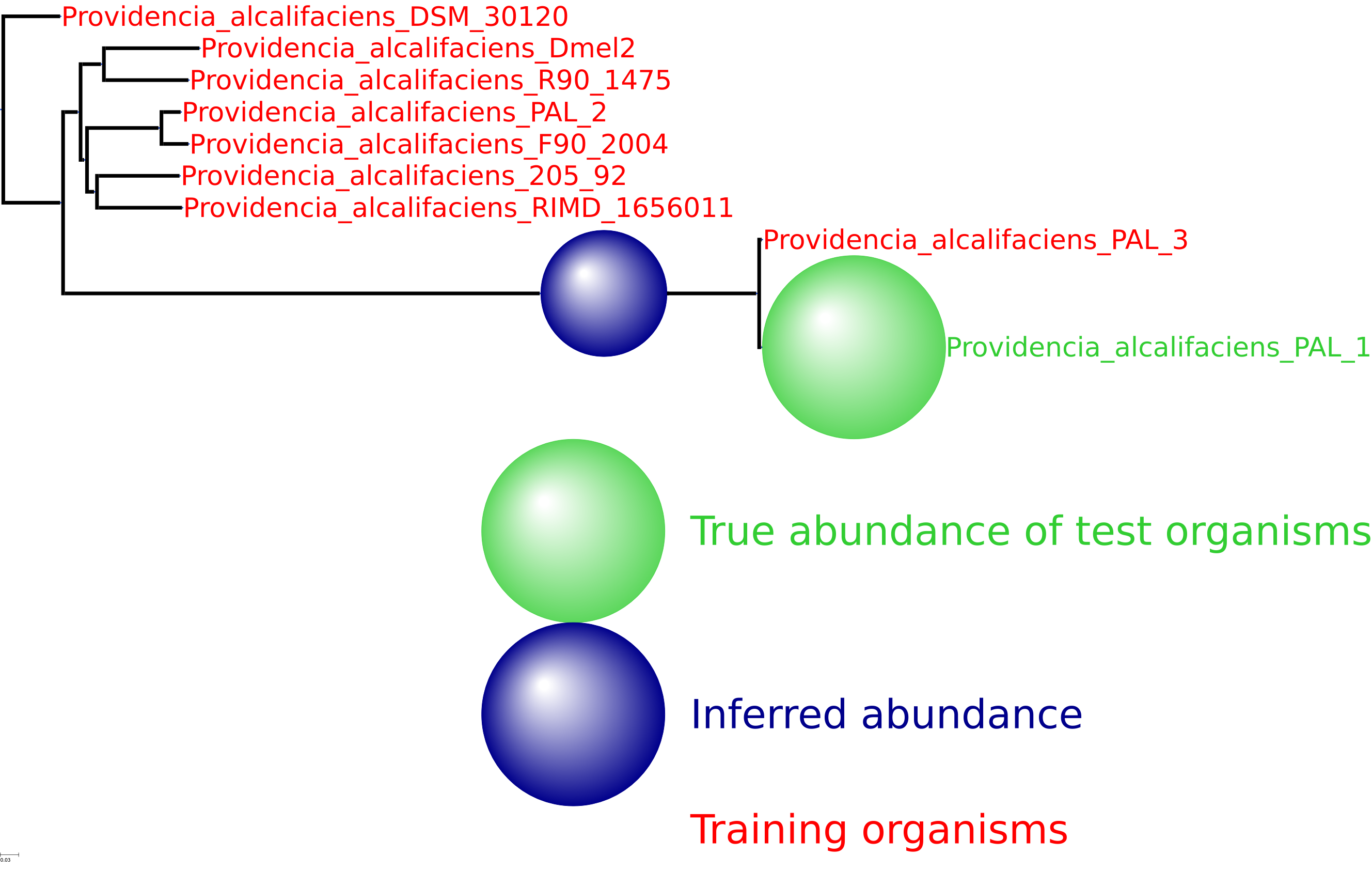}
% \caption{}
\end{minipage}
\begin{minipage}{2.5in}
(b)\begin{minipage}{2.5in}
\includegraphics[width=2.5in]{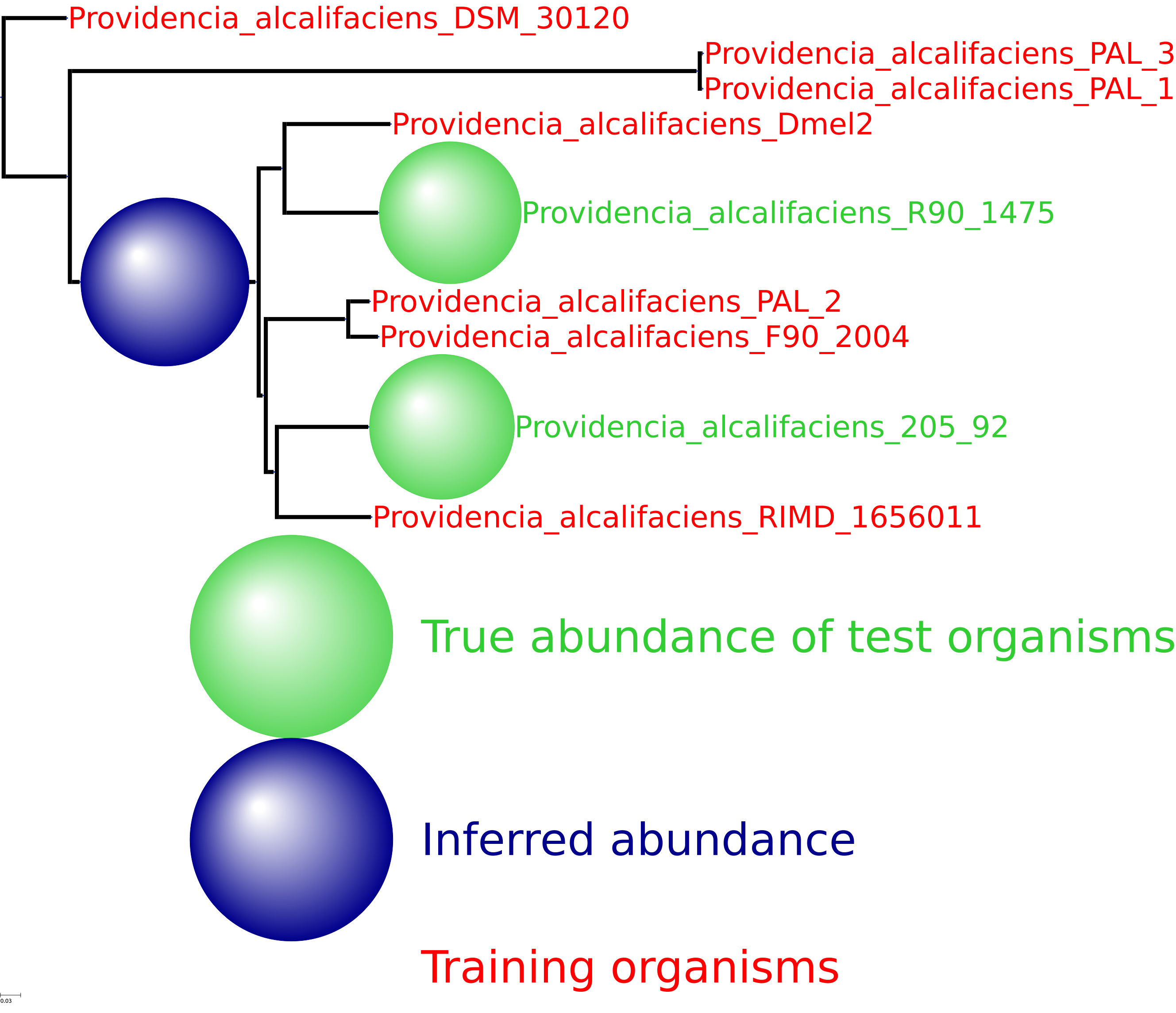}
% \caption{}
\end{minipage}
\end{minipage}
%\vspace{-5mm}
\caption{For each of the samples, a total of 50K reads from novel strains of \textit{P. alcalifaciens} were spiked into the HMP mock even community. (a) Result from training on 8 strains and testing on one novel strain. (b) Result from training 7 strains and testing on two novel strains.}
\label{fig:MorePlots}
\end{figure*}

Decreasing the number or changing the identity of the training organisms does not impede the method. In Figure \ref{fig:MorePlots}, 50K simulated reads from the species \textit{Providencia alcalifaciens} were again spiked into the HMP mock even community, and the inferred abundance is again placed optimally on the neighbor joining tree. Appendix section \ref{Appendix:AdditionalFigures} contains a variety of such figures spanning all domains of life. These results provide evidence that {\sf MetaPalette} can correctly infer the presence of organisms related to, but absent from, the training database.

\subsection{Metagenomic Soil Sample}
\begin{figure*}[h!]
\centering
\hspace{-5ex}
(a)
\begin{minipage}{3in}
\hspace{-1ex}
\includegraphics[width=3in]{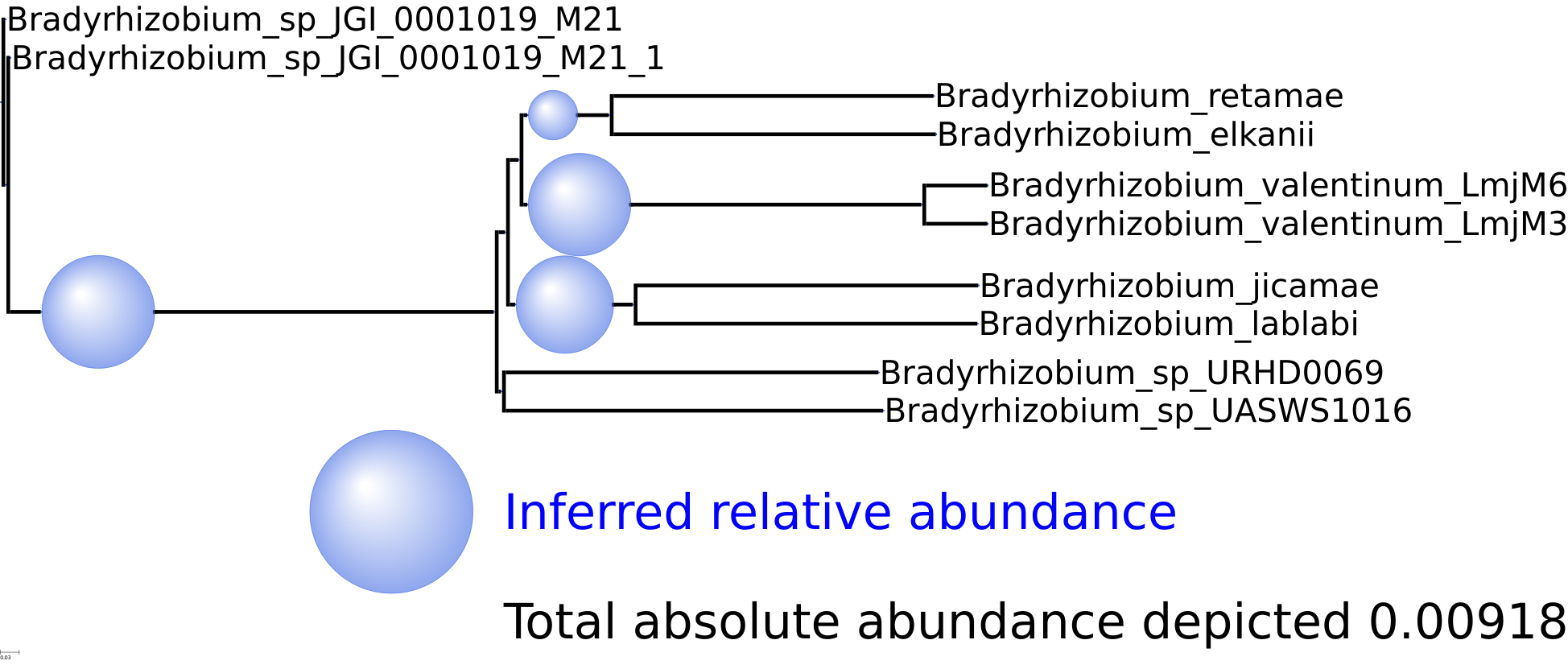}
% \caption{}
\end{minipage}
\begin{minipage}{3in}
(b)\begin{minipage}{3in}
\includegraphics[trim={5cm 1.5cm 5cm 1.5cm},clip,width=3.2in]{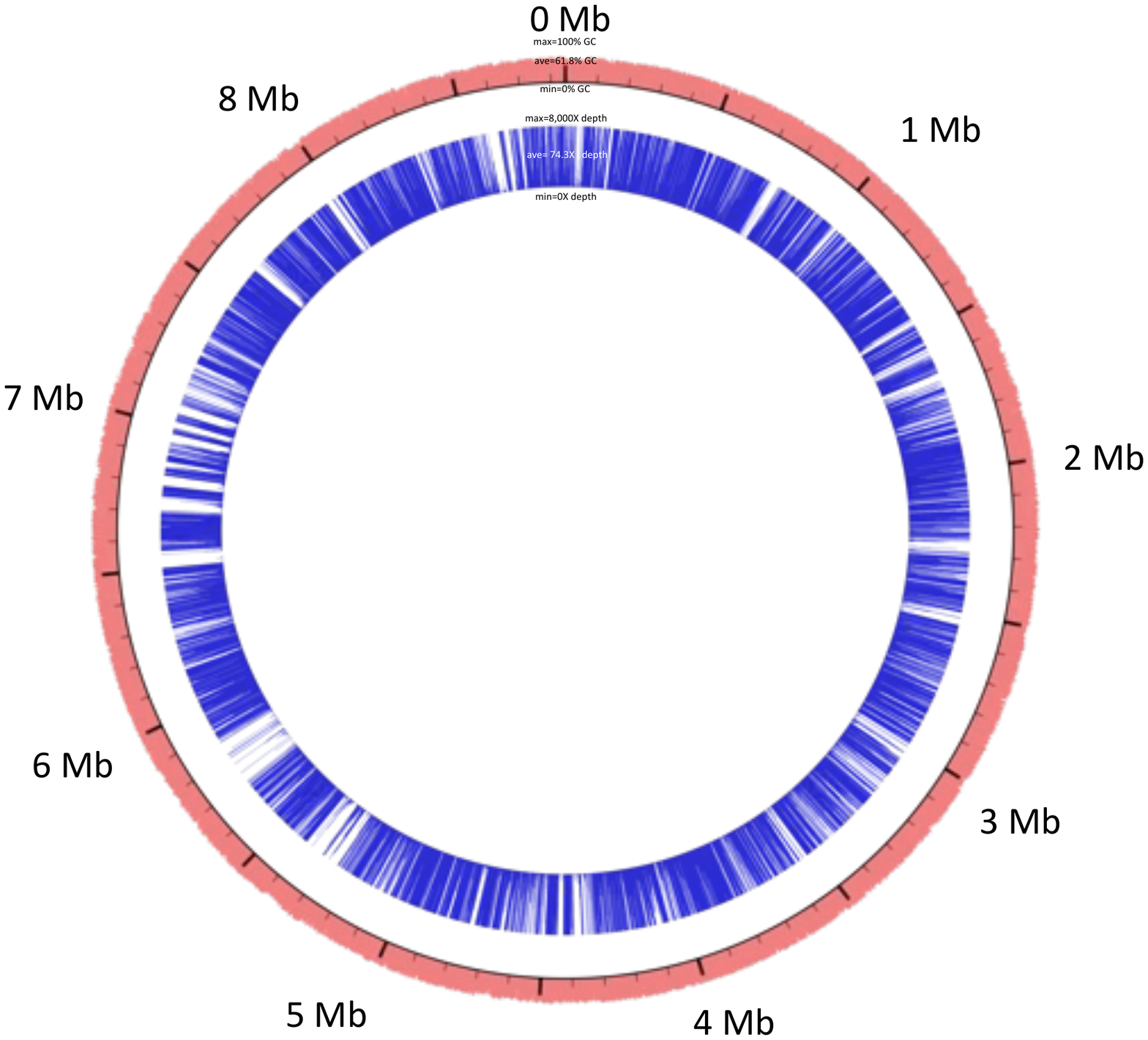}
% \caption{}
\end{minipage}
\end{minipage}
\begin{center}
(c)\begin{minipage}{4.2in}
\includegraphics[trim={4cm 2cm 5cm 2cm},clip,width=4.2in]{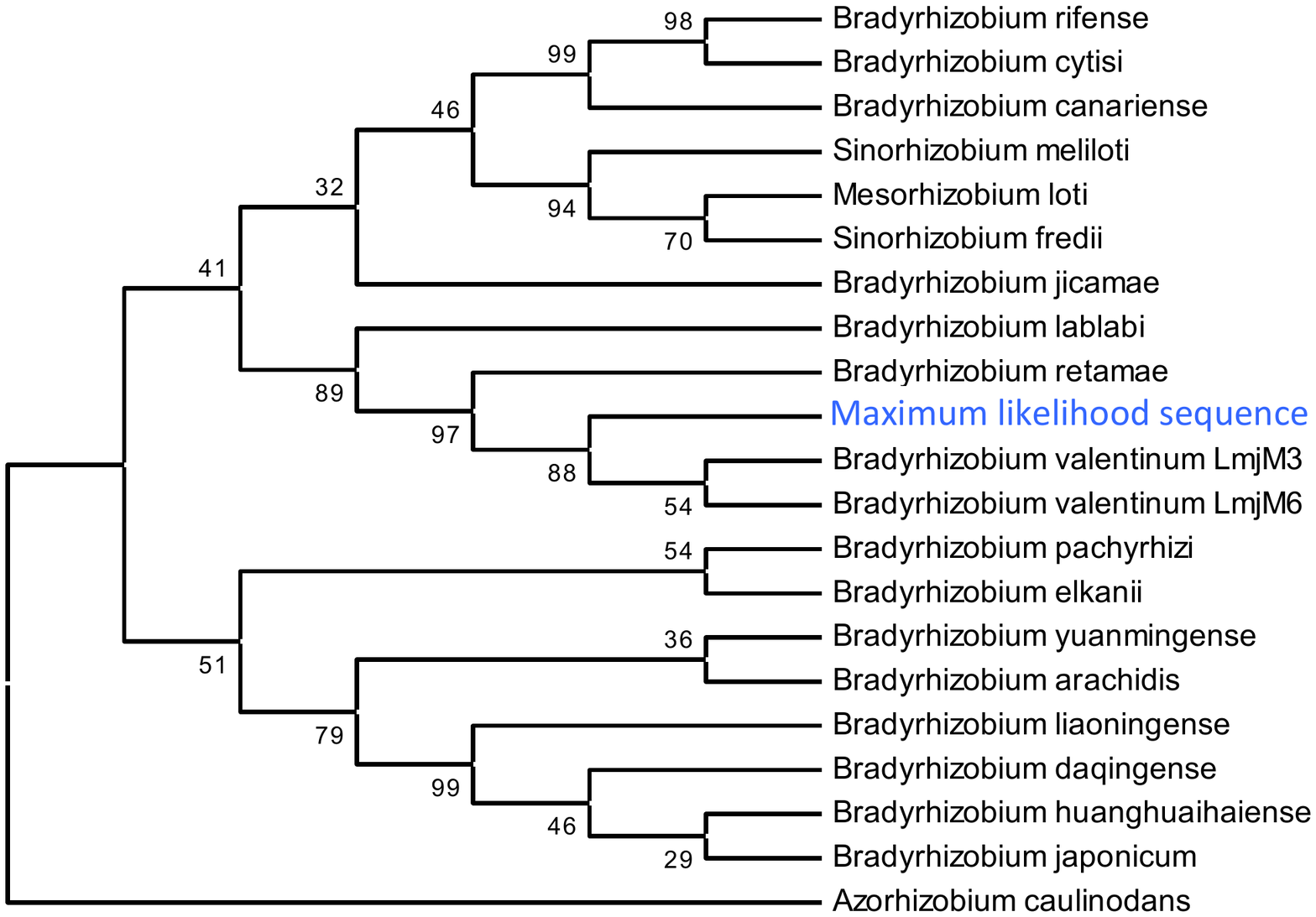}
\end{minipage}
\end{center}
%\vspace{-5ex}
\caption{(a) Subtree of the {\sf MetaPalette} output tree for the Iowa prairie metagenome using organisms from the genus \textit{Bradyrhizobium}. (b) Depth of alignment for reads from the soil metagenome that aligned to \textit{B. valentinum LmjM3}. The outer red ring shows the \%GC for \textit{B. valentinum LmjM3}, and the inner blue ring shows the alignment depth (truncated to 8,000X for ease of viewing). All contigs of the reference \textit{B. valentinum LmjM3} were concatenated in this figure. (c) Bootstrap consensus tree topology based on \textit{nifH} for 20 organisms along with the maximum likelihood sequence obtained from aligning the soil metagenome to the \textit{nifH} gene sequence of \textit{B. valentinum LmjM3}. Bootstrap values (500 replicates) are shown next to the branches. Full details regarding formation of the tree, along with a figure containing the branch lengths, are given in the Appendix section \ref{appendix:tree}. }
\label{fig:Soil}
\end{figure*}

To assess {\sf MetaPalette} on a real metagenomic sample, we utilized the Iowa prairie metagenomic sample from \cite{howe2014tackling} (corresponding to MG-RAST project ID 6377). After running {\sf MetaPalette} on a subset of this data (metagenome 4539594.3), the returned taxonomic profile predicted the presence of the genus \textit{Bradyrhizobium}. Generating the tree plot on a subset of this genus resulted in, among others, a prediction of a novel organism in the clade defined by strains of \textit{B. valentinum} (see Figure \ref{fig:Soil}(a)). To verify this, we aligned the entire soil metagenome to the reference genome of the strain \textit{B. valentinum LmjM3} using {\sf Bowtie2} with {\sf --very-sensitive-local} settings \cite{langmead2012fast} and extracted the aligned reads. Interestingly, 0.29\% of the reads aligned, while the {\sf MetaPalette} predicted abundance for this putative novel organism of interest was 0.33\%. The depth of coverage of the extracted reads is pictured in Figure \ref{fig:Soil}(b) and had a mean depth of 74.3X. 

To assess the evolutionary relatedness of this predicted organism, we utilized the \textit{B. valentinum LmjM3} \textit{nifH} gene sequence (NCBI accession KF806461) which was used in \cite{duran2014bradyrhizobium}, along with other genes, to determine the taxonomy of \textit{B. valentinum LmjM3}. Aligning the extracted reads to \textit{nifH} resulted in a mean depth of coverage of 22X. We collapsed the aligned reads (via a majority vote) in regions of coverage at least 22X and called this the maximum likelihood sequence. We then performed a multiple sequence alignment of this sequence along with the \textit{nifH} sequences of 20 other organisms closely related to \textit{B. valentinum}. The topology of the bootstrap consensus neighbor-joining tree is pictured in Figure \ref{fig:Soil}(c) and shows the maximum likelihood sequence is placed at the same location as was predicted by {\sf MetaPalette}. While this is not enough evidence to unequivocally claim the existence of a novel strain in this sample, this gives support that {\sf MetaPalette} correctly inferred the abundance and placement in Figure \ref{fig:Soil}(a) of a potentially novel strain in the clade defined by strains of \textit{B. valentinum}.

%To assess the evolutionary relatedness of this predicted organism, we utilized the \textit{B. valentinum LmjM3} \textit{nifH} gene sequence (accession KF806461)which was used in \cite{duran2014bradyrhizobium}, along with other genes, to determine the taxonomy of \textit{B. valentinum LmjM3}. Aligning the extracted reads to \textit{nifH} resulted in a mean depth of coverage of 22X. In regions of coverage greater than 22X there were 18 substitutions between the aligned reads and the reference \textit{nifH} gene sequence. Comparatively, there were no substitutions between the \textit{nifH} genes of the \textit{B. valentinum} strains, and 23 substitutions between the \textit{nifH} genes of \textit{B. valentinum LmjM3} and the closely related \textit{B. valentinum lablabi}. While this is not enough evidence to unequivocally claim the existence of a novel strain, this gives support that {\sf MetaPalette} correctly inferred the abundance and placement in Figure \ref{fig:Soil}(a) of a potentially novel strain in the clade defined by strains of \textit{B. valentinum}.

\section{Comparison to Other Metagenomic Profiling Methods}
To facilitate an objective comparison with other methods with minimal ``author bias", we utilized the same data and metrics used by other authors in a recent metagenomics methods evaluation paper \cite{lindgreen2015evaluation}. This allowed comparison to the following algorithms: {\sf CLARK} \cite{ounit2015clark}, {\sf Kraken} \cite{wood2014kraken}, {\sf OneCodex} \cite{minot2015one}, {\sf LMAT} \cite{ames2013scalable}, {\sf MG-RAST} \cite{meyer2008metagenomics}, {\sf MetaPhlAn} \cite{truong2015metaphlan2}, {\sf mOTU} \cite{sunagawa2013metagenomic}, {\sf Genometa} \cite{davenport2012genometa}, {\sf QIIME} \cite{caporaso2010qiime}, {\sf EBI} \cite{hunter2014ebi}, {\sf MetaPhyler} \cite{liu2010metaphyler}, {\sf MEGAN} \cite{huson2011integrative}, {\sf taxator-tk} \cite{droge2015taxator}, and {\sf GOTTCHA} \cite{freitas2015accurate}. 

\subsection{Training Data}
Each of the methods was trained using their default recommended databases. We trained our method using 6,914 whole genome sequences and assemblies obtained from various public repositories via {\sf RepoPhlAn} (\url{https://bitbucket.org/nsegata/repophlan}).  The training procedure for {\sf MetaPalette} on these 6,914 organisms took a total of approximately 7 hours on a 48 core server.

\subsection{Testing Data}
The testing data consisted of 6 samples, and is fully explained in the Methods section of \cite{lindgreen2015evaluation}, but we briefly summarize it here. Three replicates were formed from two different distributions of over 900 different genomes spanning the tree of life (including Eukaryote genomes). Included in each test sample were shuffled/randomized genomes (not meant to be assigned to any known taxa) as well as sequences from the genome of \textit{Leptospira interrogans} that were evolved using {\sf Rose} \cite{stoye1998rose} to simulate novelty. Error profiles were based on those of 6 real soil metagenomic samples sequenced using an Illumina HiSeq 2000. Each of the resulting test samples contains between 27 and 37 million read pairs.

\subsection{Error Metrics}
We utilized the same divergence error metric as \cite{lindgreen2015evaluation}, that is, for $x_i$ representing the true frequency of taxa $i$ in the sample, and $x^*_i$ representing the predicted frequency for a given method of taxa $i$, 
$$
{\rm Divergence} = \sum_i \log_2\left(\frac{x^*_i}{x_i}\right)
$$
where the summation is over those indices such that $x_i>0$ and $x^*_i>0$. Since this error metric does not take into consideration the number of spurious assignments (that is, taxa predicted by a method to be in a sample, but not actually present), we also use the number of false positives at a given taxonomic rank:
$$
{\rm FP} = |\{i:\ x^*_i>0\ {\rm and}\ x_i=0\}|.
$$

\subsection{Comparison Results}
\begin{figure*}[h!]
\centering
(a)
\begin{minipage}{3in}
\includegraphics[width=3in]{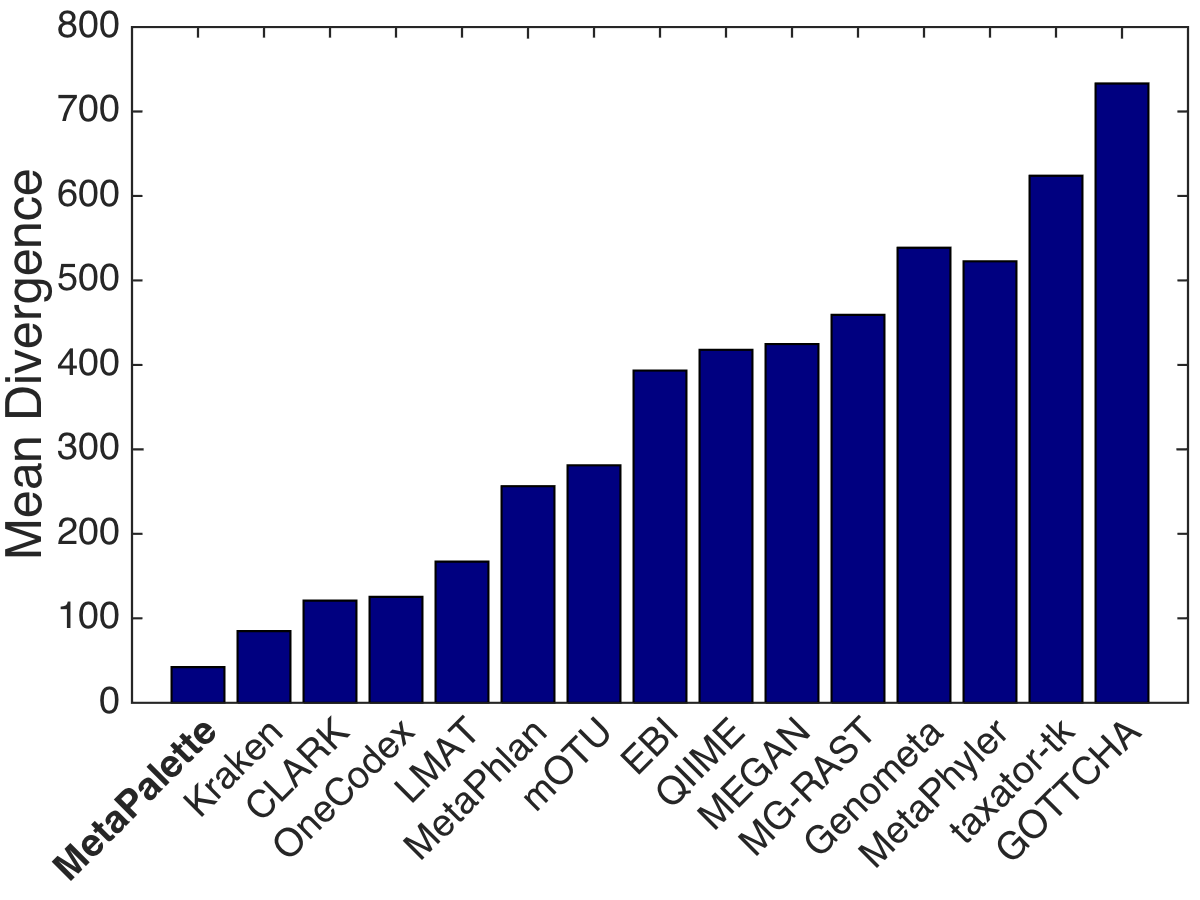}
% \caption{}
\end{minipage}
\begin{minipage}{3in}
(b)\begin{minipage}{3in}
\includegraphics[width=3in]{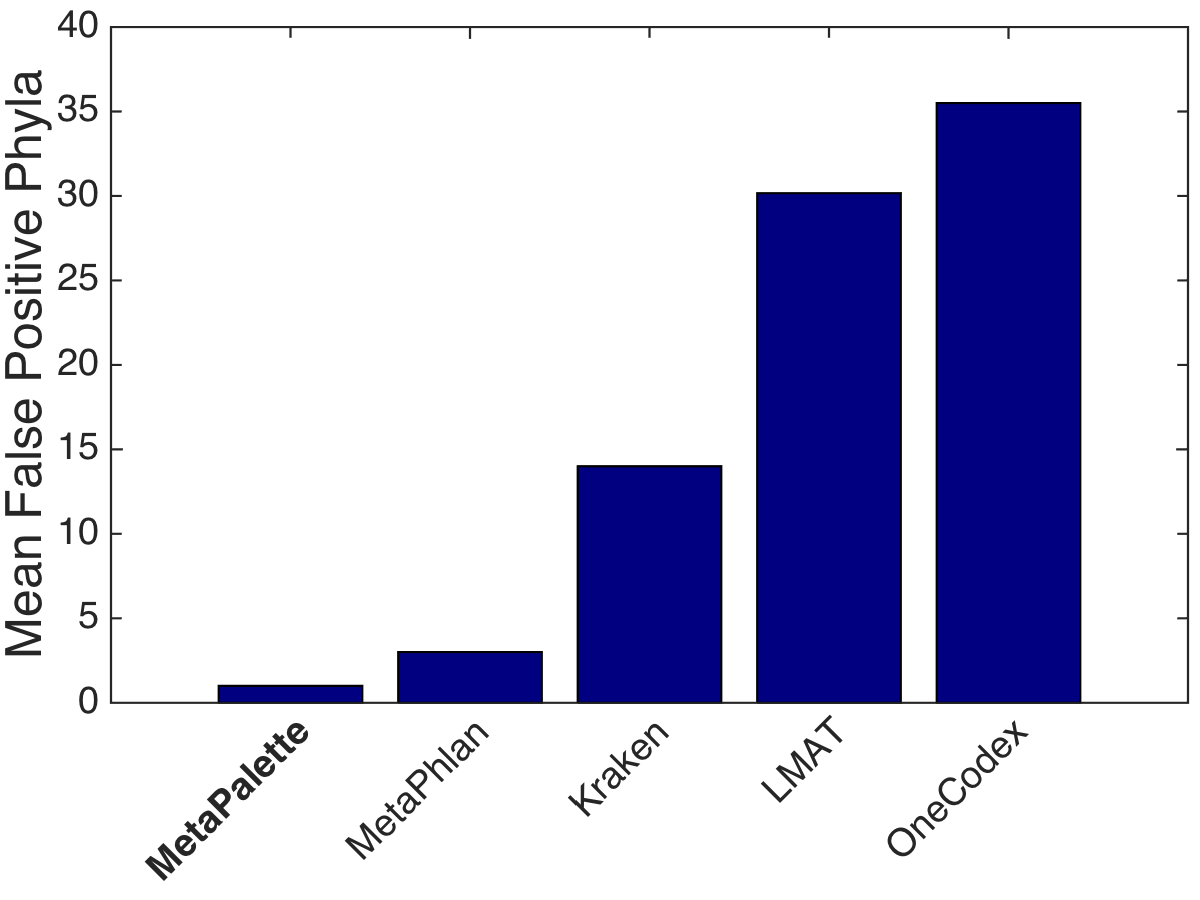}
% \caption{}
\end{minipage}
\end{minipage}
%\vspace{-5mm}
\caption{Plot of performance metrics for all methods averaged over all test samples. Smaller values indicate better performance. (a) Divergence error metric at the genus level. (b) Number of false positive phyla.}
\label{fig:TestResults}
\end{figure*}

Each method was run using the default parameters. For each method, we averaged the divergence error metric over all the test samples at the genus level (see Figure \ref{fig:TestResults}(a)). Furthermore, we selected a number of the more accurate methods and averaged the number of false positives over all the test samples at the phylum level (see Figure \ref{fig:TestResults}(b)). These two figures clearly show the competitive nature of {\sf MetaPalette} as it has the lowest error in both metrics. However, when comparing to other methods, one should be careful of their intended use. For example, {\sf taxator-tk} is intended to be used on an assembled metagenome (and here unassembled reads were used), and {\sf QIIME} only uses the 16S rRNA sequences in a sample. Furthermore, most of these methods assign individual reads and then summarize this to obtain a taxonomic profile, while our method only profiles the entire sample and returns relative proportions of organisms.

\begin{figure}[h!]
\includegraphics[width=3in]{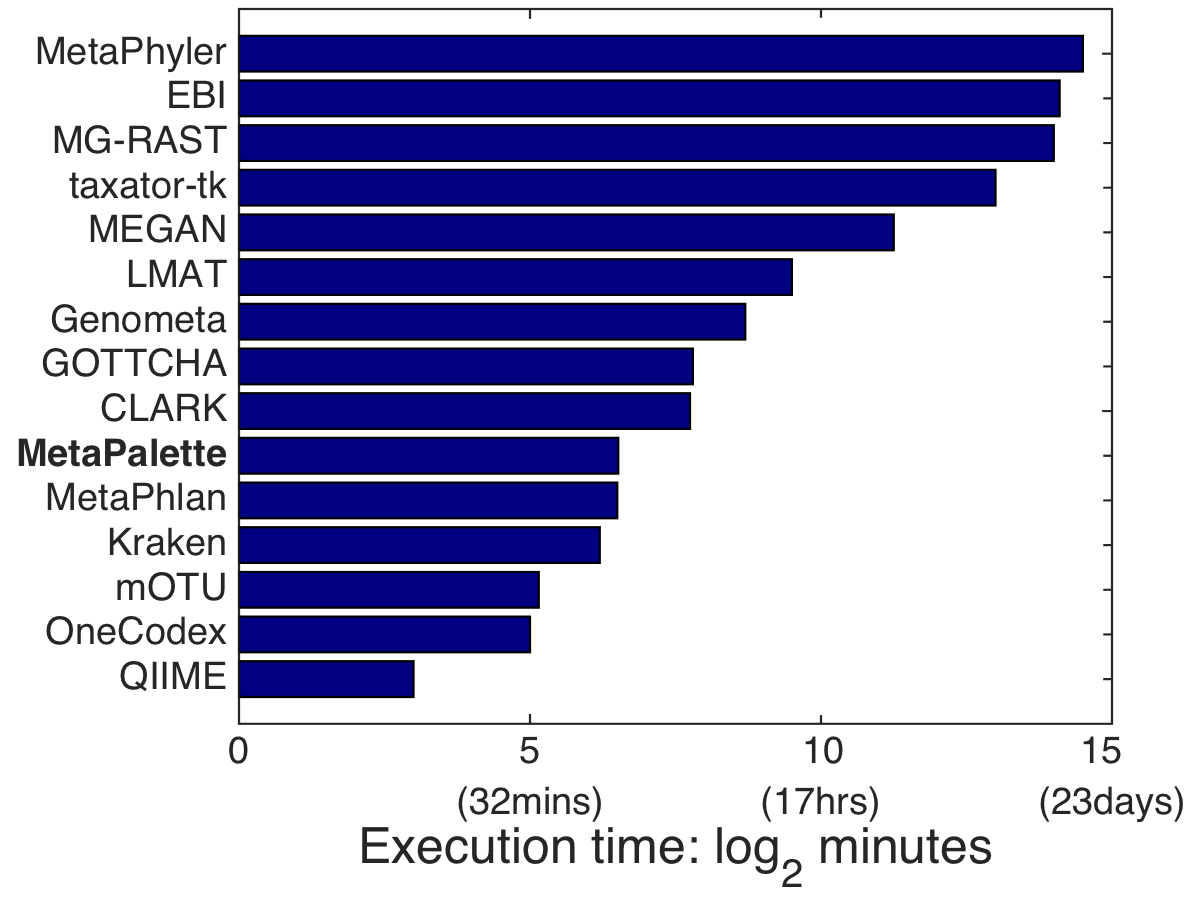}
\caption{Mean execution time of each method averaged over all 6 test samples.}
\label{fig:Timing}
\end{figure}

Figure \ref{fig:Timing} shows the timing of each of the methods (on a log scale, obtained from \cite{lindgreen2015evaluation}) further showing the competitive nature of {\sf MetaPalette}.

%\label{section:OtherMethods}
%We compare this method to other metagenomic profiling methods by creating a number of realistic simulated test datasets and employing a variety of assessment metrics. The other methods we compare to are {\sf MetaPhlAn2} \cite{truong2015metaphlan2}, {\sf mOTU} \cite{sunagawa2013metagenomic}, {\sf TIPP} \cite{nguyen2014tipp}, {\sf Kraken} \cite{wood2014kraken}, and {\sf Clark} \cite{ounit2015clark}. These methods constitute some of the most popular community profilers (in the case of {\sf MetaPhlAn2}) and the most accurate (in the case of {\sf Kraken} and {\sf Clark}) \cite{That review paper}. 

%\subsubsection{Training Data}
%We used {\sf RepoPhlAn} to obtain $\sim 30,000$ whole microbial genome sequences and randomly selected $1,000$ to train the common $k$-mer method. Each of the other methods was employed using the default training data and parameter settings.

\vspace{1ex}
\section{Software and Pre-trained Data}
\subsection{Software}
The source code for {\sf MetaPalette}, along with installation instructions and directions, is accessible at \url{https://github.com/dkoslicki/MetaPalette}. {\sf MetaPalette} is written primarily in python and accepts input reads in uncompressed fasta or fastq format, as well as compressed fasta/fastq using {\sf bzip2} and {\sf gzip}. For fastq input, optional parameters can be given to specify only counting $k$-mers above a certain quality score (Phred) thereby attenuating the negative impact of sequencing error in the correct inference of relative abundances. The output taxonomic profile is compliant with the Bioboxes profiling format version 0.9 found at \url{https://github.com/bioboxes/rfc/tree/master/data-format}. Python scripts are also included to aid in downloading data, forming custom databases, and creating the appropriate taxonomy files. 

To facilitate cross-platform usability, a Docker \cite{merkel2014docker} container has been created and is accessible at: \url{https://hub.docker.com/r/dkoslicki/metapalette} with an accompanying docker file at: \url{https://github.com/dkoslicki/MetaPalette/blob/master/Docker/Dockerfile}.

If users wish to use {\sf MetaPalette} but lack computational resources, they may utilize the {\sf Galaxy} \cite{giardine2005galaxy,goecks2010galaxy,blankenberg2010galaxy} server located at: \url{http://math-galaxy.cgrb.oregonstate.edu/}.

A preliminary version of this software was submitted to the Critical Assessment of Metagenomic Interpretation (CAMI: \url{http://www.cami-challenge.org/}) under the name of {\sf CommonKmers}. However, since significant changes have been made since that point, we strongly recommend using the current {\sf MetaPalette} software instead.

\subsection{Pre-trained Data}
To decrease computational burden, pre-trained databases are accessible at \url{http://files.cgrb.oregonstate.edu/Koslicki_Lab/MetaPalette}. Databases and accompanying taxonomies have been included for archaea (666 organisms, 1.7 gigabytes uncompressed), bacteria (15,147 organisms, 60GB), eukaryota (1,307 organisms, 41GB), and viruses (4,798 organisms, 0.6GB). All organisms were obtained via {\sf RepoPhlAn}.

The 6,914 organism database used for the comparison to other profiling methods is accessible at \url{http://files.cgrb.oregonstate.edu/Koslicki_Lab/MetaPallete/Comparison}.

\section{Conclusion}
We have described a fast, flexible and accurate method for estimating the taxonomic composition of organisms which is based on reconstructing a $k$-mer-based profile of a sample. Each reference organism has a $k$-mer ``palette" and we fit the sample as a mixture of different palettes, both of the reference organisms and organisms absent from the training data at varying degrees of relatedness to the training database. Our approach is in part inspired by chromosome painting method used to deduce fine-scale population structure in human genetics \cite{leslie2015fine,hellenthal2014genetic} which is also based on mixture modeling of palettes. A particular advantage of {\sf MetaPalette} over other metagenomic profiling methods is that {\sf MetaPalette} provides an indication of how related the organisms in a given sample are to the closest matching organisms of the training database, whether they are within the same species or distantly related organisms from the same phyla. 

Furthermore, the standard approach to summarizing composition information has been to place organisms at different taxonomic levels. We produce a standard taxonomic profile which we have shown to be more accurate than that produced by other methods. This fixed rank approach is sensible at the genus level and above but omits fine-scale information. Hence, for branches of the tree of life that are well represented in the training database, we can also output a phylogenetic tree giving detailed information on how the sampled taxa relate to the organisms in the training database (Figures \ref{fig:TwoAbsent}-\ref{fig:Soil} and Appendix section \ref{Appendix:AdditionalFigures}).

%A particular advantage of our method compared to most of the competition that it provides an indication of how related the organisms are to the closest matching organisms in the reference set, whether they are within the same species or distantly related organisms from the same phyla. 

%Paragraph on what data are needed and procedure. 

%The standard approach to summarizing composition information has been to bin organisms at different taxonomic levels. This approach is sensible at the genus level and above but omits fine-scale information. We produce a standard taxonomic binning which we have shown is more accurate than that produced by other software. For branches of the tree of life that are well represented in the training dataset, we can also output a phylogenetic tree giving detailed information on how the sampled taxa relate to the organisms in the training dataset(Figures X and Y). 

For many applications, it is of interest to understand which individual reads belong to which organisms \cite{sharpton2014introduction}. A principled approach to this problem is to first estimate the overall composition of the sample, using {\sf MetaPalette} or an equivalent, and then to assign individual reads conditional on the overall assignment. This represents a promising avenue for future methodological development.  

\section{Acknowledgements} 
The authors would like to thank the Isaac Newton Institute at Cambridge University for their hospitality during the program on metagnomics. This project was conceived while both authors were attending this program. Daniel Falush is supported by a Medical Research Fellow fellowship as part of the CLIMB Consortium for Medical Microbiology.

The authors would also like to thank Nam Nguyen and Daniel Alemany who both contributed to this project in its preliminary stages.

%\newpage
\appendix

\section{Technical Details of the Common $K$-mer Method}
\label{appendix:method}
\subsection{Mathematical Formulation}
\label{appendix:MathFormulation}
We include here all rigorous mathematical definitions of the quantities discussed in the main text.

Given the alphabet $\mathscr{A}=\{A,C,T,G\}$, let $\mathscr{A}^n$ denote the set of all words $v$ of length $|v|=n$ on $\mathscr{A}$, and let $\mathscr{A}^*=\bigcup_{n\geq 0} \mathscr{A}^n$ be the set of all finite words on $\mathscr{A}$. {Hence words containing non-$ACTG$ characters are ignored.} Let $D=\{g_1,\dots,g_M\}$ be a database of genomic sequences $g_j\in \mathscr{A}^*$ and let $S=\{s_1,\dots,s_N\}$ be a set of sample sequences (the reads to be classified). For notational simplicity, assume that the read length is fixed: for all $t$, $|s_t|=r$. Fix a $k$-mer size and endow $\mathscr{A}^k=\{v_1,\dots,v_{4^k}\}$ with the lexicographic order. Let ${\rm occ}_v(w)$ represent the number of occurrences (with overlap) of the subword $v$ in the word $w$. That is, for $w,v\in \mathscr{A}^n$, let
\begin{align}
{\rm occ}_v(w)=|\{j:w_j w_{j+1}\cdots w_{j+|v|-1}=v\}|.
\end{align}
For a fixed $k$-mer size, and two genomes $g_i$ and $g_j$ we calculate the number of $k$-mers in genome $j$ common to both $g_i$ and $g_j$. That is, the $(i,j)^{\rm th}$ entry of the common $k$-mer training matrix $A^{(k)}$ is:
\begin{align}
\label{eqn:CommonKmerMatrix}
A^{(k)}_{i,j} = \sum_{w\in \SW_{g_i}(k)\bigcap \SW_{g_j}(k)} \frac{\occ_w(g_j)}{|g_j|-k+1}.
\end{align}
Refer to the entries of the common $k$-mer matrix as ${\rm pckm}_k(g_i,g_j) = A^{(k)}_{i,j}$.
Let $s_i \subset g_j$ denote the relationship that read $s_k$ was derived from genome $g_j$. We represent the taxonomic profile of the sample $S$ by the probability vector $x$:
\begin{align}
x_j = \frac{1}{N} \sum_{t=1}^N \mathbbm{1}_{\{s_t \subset g_j\}}
\end{align}
where $\mathbbm{1}$ is the indicator function. Now let the measurement vector $y$ be given by the probability vector
\begin{align}
y^{(k)}_i= \sum_{w\in \SW_{g_i}(k)\bigcap \SW_S(k)} \occ_w(S) / \sum_{w\in \SW_S(k)} \occ_w(S).
\end{align}
We assume that the reads $s_t$ are uniformly randomly selected from the genomes $g_j$. Then for $w\in \AAA^k$, let $\mathbb{P}(w|g_j)$ be the probability that $k$-mer $w$ is found in genome $g_j$. Then we have that the proportion of $k$-mers $w$ in the sample is similar to the proportion of the appearance of $w$ in the genomes $g_j$ when weighted by the relative abundance of the genomes $g_i$ in the sample:
\begin{align}
\label{eqn:ProbEst}
\frac{\sum_{t=1}^N \occ_w(s_t)}{\sum_{w\in \SW_S(k)} \occ_w(S)} &\approx \frac{1}{N}\sum_{j=1}^M \mathbb{P}(w|g_j) \sum_{t=1}^N \mathbbm{1}_{\{s_t\subset g_j\}}\\
& = \frac{1}{N}\sum_{j=1}^M \frac{\occ_{g_j}(w)}{|g_j|-n+1} \sum_{t=1}^N \mathbbm{1}_{\{s_t\subset g_j\}}
\end{align}
We then calculate
\begin{align}
y^{(k)}_i &= \frac{1}{\sum_{w\in \SW_S(k)} \occ_w(S)}\sum_{w\in \SW_{g_i}(k)\bigcap \SW_S(k)} \occ_w(S)\\
%&=\frac{1}{\sum_{w\in \SW_S(k)}\occ_w(S)}\sum_{w\in \SW_{d_i}(k)\bigcap \SW_S(k)} \occ_w(S)\\
& =\frac{1}{\sum_{w\in \SW_S(k)}\occ_w(S)} \sum_{w\in \SW_{g_i}(k)\bigcap \SW_S(k)}\quad \sum_{t=1}^N \occ_w(s_t) \\ 
&\approx \frac{1}{N}\sum_{w\in \SW_{g_i}(k)\bigcap \SW_S(k)}\quad \sum_{j=1}^M \frac{\occ_{g_j}(w)}{|g_j|-k+1}\quad \sum_{t=1}^N \mathbbm{1}_{\{s_t\subset g_j\}}\\  
&=\frac{1}{N}\sum_{j=1}^M\quad  \sum_{w\in \SW_{g_i}(k)\bigcap \SW_{g_j}(k) \bigcap \SW_S(k)} \frac{\occ_{g_j}(w)}{|g_j|-k+1}\quad \sum_{t=1}^N \mathbbm{1}_{\{s_t\subset g_j\}}\label{whatever} \\
&\approx \frac{1}{N}\sum_{j=1}^M \quad \sum_{w\in \SW_{g_i}(k)\bigcap \SW_{g_j}(k)} \frac{\occ_{g_j}(w)}{|g_j|-k+1}\quad \sum_{t=1}^N \mathbbm{1}_{\{s_t\subset g_j\}}\label{intersectline} \\
&= \sum_{j=1}^M A^{(k)}_{i,j} x_j = \left( A^{(k)} x \right)_i
\end{align}
Line \eqref{whatever} is justified since if $w \notin \SW_{g_j}(k)$ then $\occ_{g_i}(w)=0$. For computational reasons, we make the assumption  in line \eqref{intersectline} that $\SW_{g_i}(k)\bigcap \SW_{g_j}(k) \bigcap \SW_S(k) = \SW_{g_i}(k)\bigcap \SW_{g_j}(k)$. However, this assumption can be mitigated by adding hypothetical organisms (see section \ref{section:AppendixHypOrg}).
 Our assumptions imply that
\begin{align}
\label{eqn:matrixeqn}
A^{(k)}x\approx y^{(k)}.
\end{align}
We will try to recover the vector $x$ satisfying $x_j\geq 0$ for all $j=1,\dots,M$ from equation \eqref{eqn:matrixeqn}. 

\subsection{Further Improvements}
A few further improvements are possible, but not pursued here. Namely, we could use just the $k$-mers that are actually in the sample to form the training matrix. I.e use $\SW_{g_i}(k)\bigcap \SW_{g_j}(k) \bigcap \SW_S(k)$ in the formation of $A^{(k)}$:
$$
A^{(k)}_{i,j} = \sum_{w\in \SW_{g_i}(k)\bigcap \SW_{g_j}(k)\bigcap \SW_S(k)} \frac{\occ_w(g_j)}{|g_j|-k+1}.
$$
The disadvantage of this is that the (slow) training step would need to be re-run for each sample.

For a second improvement, we could make the approximation in \eqref{eqn:ProbEst} more delicate by incorporating the coverage:
\begin{align}
\frac{\sum_{t=1}^N \occ_w(s_t)}{\sum_{w\in \SW_S(k)} \occ_w(S)} &\approx \frac{1}{N}\sum_{j=1}^M \mathbb{P}(w|g_j) \frac{\sum_{w\in \SW_{g_j}(k) \bigcap \SW_S(k)}\occ_w(g_j)}{\sum_{w\in \SW_{g_j}(k)} \occ_w(g_j)} \sum_{t=1}^N \mathbbm{1}_{\{s_t\subset g_j\}}\\
& = \frac{1}{N}\sum_{j=1}^M \frac{\occ_{g_j}(w)}{|g_j|-n+1} \frac{\sum_{w\in \SW_{g_j}(k) \bigcap \SW_S(k)}\occ_w(g_j)}{|g_j|-k+1}\sum_{t=1}^N \mathbbm{1}_{\{s_t\subset g_j\}}.
\end{align}
So $A^{(k)}$ would have the form:
\begin{align}
A^{(k)}_{i,j} = \frac{\sum_{w\in \SW_{g_j}(k) \bigcap \SW_S(k)}\occ_w(g_j)}{|g_j|-k+1} \sum_{w\in \SW_{g_i}(k)\bigcap \SW_{g_j}(k)\bigcap \SW_S(k)} \frac{\occ_w(g_j)}{|g_j|-k+1}. \label{eq:coverage}
\end{align}
This effectively multiplies the columns $j$ of $A^{(k)}$ by the $\%$ coverage of genome $j$.
Lastly, in \eqref{eq:coverage}, we could put a weighting factor that represents how unique a $k$-mer is to the genome in question. This would down-weight $k$-mers shared among many diverse genomes and up-weight those unique to certain strains/species/genera/etc.

\subsection{Hypothetical Organisms}
\label{section:AppendixHypOrg}
To simulate an organism that is, say, $90\%$ related to a database genome $g_i$, we augment the common $k$-mer matrix $A^{(k)}$ with a column derived by rounding down the entries of the column vector $\{A^{(k)}_{i,j}\}_{i=1,\dots, M}$ that are above $90\%$. 
Two $k$-mer sizes are needed to form the hypothetical organism common $k$-mer matrices. For the first $k$-mer size, $k_1$, we define $A^{(k_1),h}$ for a fixed number of hypothetical bins $h\in \{0.9, 0.8,  \dots, 0.1\}$ where
\begin{align}
\label{eqn:A^{(k_1),h}}
A^{(k_1),h}_{i,j} = \max \left( h, \sum_{w\in \SW_{g_i}(k_1)\bigcap \SW_{g_j}(k_1)} \frac{\occ_w(g_j)}{|g_j|-n+1}\right).
\end{align}
For the second $k$-mer size, $k_2$, using the polynomial $p(x)=-.5141x.^3+1.0932x.^2+0.3824x$, we define
\begin{align}
\label{eqn:A^{(k_2),h}}
A^{(k_2),h}_{i,j} = \max \left( h, p\left(\sum_{w\in \SW_{g_i}(k_2)\bigcap \SW_{g_j}(k_2)} \frac{\occ_w(g_j)}{|g_j|-n+1}\right)\right).
\end{align}
Instead of thresholding, as we did here, one can imagine other scalings obtained from studying the relationship between a given taxonomy and the common $k$-mer matrix $A^{(k)}$. In particular, to deal with differing rates of evolution in the tree of life, a fruitful area of future investigation would be to modify the polynomial $p$ depending on the taxonomy of the organisms under consideration.

\subsection{Optimization Procedure}
\label{section:OptimizationProcedure}
We choose two $k$-mer sizes to be $k=30$ and $k=50$, as this seems to give a good trade-off between reconstruction fidelity and computational performance. We then collect the common $k$-mer matrix and hypothetical matrices block-wise into the $2|D| \times 10|D|$ size matrix
\begin{align}
\label{eqn:A}
A = \left[\begin{array}{cccc}
 A^{(30)}, &  A^{(30),0.9}, & \dots, & A^{(30),0.1}\\
 A^{(50)}, &  A^{(50),0.9}, & \dots, & A^{(50),0.1}
 \end{array} \right].
\end{align}
Collect also the $k$-mer sample vectors $y^{(k)}$:
\begin{align}
y = \left[\begin{array}{c}
y^{(30)}\\
y^{(50)}
\end{array}\right].
\end{align}
The problem at hand is then to reconstruct the phylogenetic profile $x$ by solving the linear system
\begin{equation}\label{eqn:Ax=y}
Ax=y.
\end{equation}

Equation \eqref{eqn:Ax=y} is solved by using a sparsity-promoting optimization procedure motivated by techniques used in the compressive sensing literature. Sparsity is emphasized due to the inclusion of the hypothetical organisms, as well as the reasonable assumption that relatively few organisms from the database $D$ are actually present in the given sample. We use a variant of nonnegative basis pursuit denoising which reduces to a nonnegative least squares problem \cite{FoucartKoslicki2014,chen1998}.
We aim to solve
\begin{align}
\label{eq:L1min}
x^* = \argmin_{ z}\  ||z||_1 \quad {\rm subject\ to}\ { A} { z}= {y}, { z}\geq 0.
 \tag{$\ell_1$-min}
\end{align}
This optimization procedure has the advantage of being transformed into a nonnegative least squares problem. Indeed as $\lambda \rightarrow \infty$, we can regularize \eqref{eq:L1min} as
\begin{align}
\label{eq:NNREG}
x^* = \argmin_{z}\ || z ||_1^2 + \lambda^2 ||{A}{ z}-{y}||_2^2 \quad {\rm subject\ to}\ { z}\geq 0.
 \tag{NNREG}
\end{align}
This reduces to a nonnegative least squares problem by defining
$$
\tilde{{ A}}\vcentcolon=
\left[\begin{tabular}{c} $1 \cdots 1$\\
\hline $\lambda \, { A}$
\end{tabular}\right], \quad \quad
\tilde{y}\vcentcolon= \left[ \begin{tabular}{c} 0\\
\hline $\lambda \, {y}$
\end{tabular} \right].
$$
So \eqref{eq:NNREG} is equivalent to the nonnegative least squares problem
\begin{align*}
x^* = \argmin_{ z}\ ||\tilde{A}{ z}-\tilde{{ y}}||_2^2 \quad {\rm subject\ to}\ { z}\geq 0.
\end{align*}
This can be solved efficiently by using the Lawson--Hanson algorithm \cite{lawson1974solving}. We use the value $\lambda=200$ throughout as this value gives a good trade-off between sparsity and accuracy of fit of the $k$-mer counts.

%As the matrix $A$ consists of the vertical concatenation of the common $30$-mer and $50$-mer matrices, we ensure that the $30$-mers or $50$-mers are not preferentially being favored in the optimization procedure by first solving the procedure \eqref{eq:NNREG} for the $30$-mer and $50$-mer matrices, and then weight \eqref{eq:NNREG} by the ratio of these residuals. {\color{red} Say this better}.

\subsection{Inferring Taxonomy}
\label{Appendix:InferringTaxonomy}
Since the reconstructed vector $x$ may have non-zero entries corresponding to a hypothetical bin, we need to develop a method to map from a hypothetical bin to a specific taxonomic rank. A na\"{i}ve approach would be to assign a fixed taxonomic rank to each hypothetical bin (call this the \textit{fixed rank} method). For example, all non-zero entries of $x$ corresponding to $A^{(k)}$ would be assigned to the strain level, all non-zero entries of $x$ corresponding to $A^{(k),0.9}$ would be assigned to the species level, etc.

We take a more biologically informed approach:  we take the least common ancestor (LCA) taxa between a hypothetical organism and a nearby organism in the database $D$: if $x_i > 0$ corresponds to the hypothetical bin $h$, find an organism $g_j$ such that $|A^{(k)}_{i,j}-h|<\delta$ for some threshold $\delta$. In the output taxonomic profile, we assign $x_i$ to the lowest taxonomic rank common to organisms $g_i$ and $g_j$. For the output strain variation figures, we assign the abundance $x_i$ relative to the least common ancestor of $g_i$ and $g_j$ (above the LCA if $h< A^{(k)}_{i,j}$ and below the LCA if $h> A^{(k)}_{i,j}$).

For the output taxonomic profile, a hybrid of the fixed rank and LCA approaches can increase sensitivity or specificity. We thus include three options: the default option is the LCA approach, while the sensitive and specific options are  varying hybrids of the two methods.

%\subsection{Reconstruction Metrics}
% We denote the \textit{actual} and \textit{predicted} concentrations of the bacteria as probability vectors $x$ and $x^*$ respectively.
%The reconstruction metric primarily employed herein is the $\ell_1$ distance between $x$ and $x^*$: $||x-x^*||_{\ell_1}$. This quantity takes values between 0 and 2 (with perfect reconstruction being $||x-x^*||_{\ell_1}=0$) and is commonly referred to as ``total error'' (as it is the total of the absolute errors). Note that $0\leq ||x-x^*||_{\ell_1} \leq 2$.
%The term \textit{reconstruction fidelity} will be used to communicate generically how well $x^*$ approximates $x$.

%We also used binary classification metrics: we say that organism $i$ is a true positive if both $x_i>0$ and $x^*_i>0$, a false positive if $x_i=0$ and $x^*_i>0$, a false negative if $x_i>0$ and $x^*_i=0$, and a true negative if $x_i=0$ and $x^*_i=0$ . Note this does not take into account the relative abundance, just the presence or absence of an organism. This allows us to also assess the precision, sensitivity, specificity, and accuracy of a reconstruction.

\section{Sequence analysis details}
\label{appendix:tree}
In Figure \ref{fig:Soil}(c), the evolutionary history was inferred using the Neighbor-Joining method \cite{saitou1987neighbor}. The bootstrap consensus tree inferred from 500 replicates is taken to represent the evolutionary history of the taxa analyzed \cite{felsenstein1985confidence}. Branches corresponding to partitions reproduced in less than 50\% bootstrap replicates are collapsed. The percentage of replicate trees in which the associated taxa clustered together in the bootstrap test (500 replicates) are shown next to the branches. The analysis involved 21 nucleotide sequences. All positions with less than 95\% site coverage were eliminated. That is, fewer than 5\% alignment gaps, missing data, and ambiguous bases were allowed at any position. There were a total of 652 positions in the final dataset. Evolutionary analyses were conducted in MEGA6 \cite{tamura2013mega6}.

\begin{figure}[hb!]
    \centering
    \includegraphics[trim={5cm 2cm 5cm 4cm},clip,width=5in]{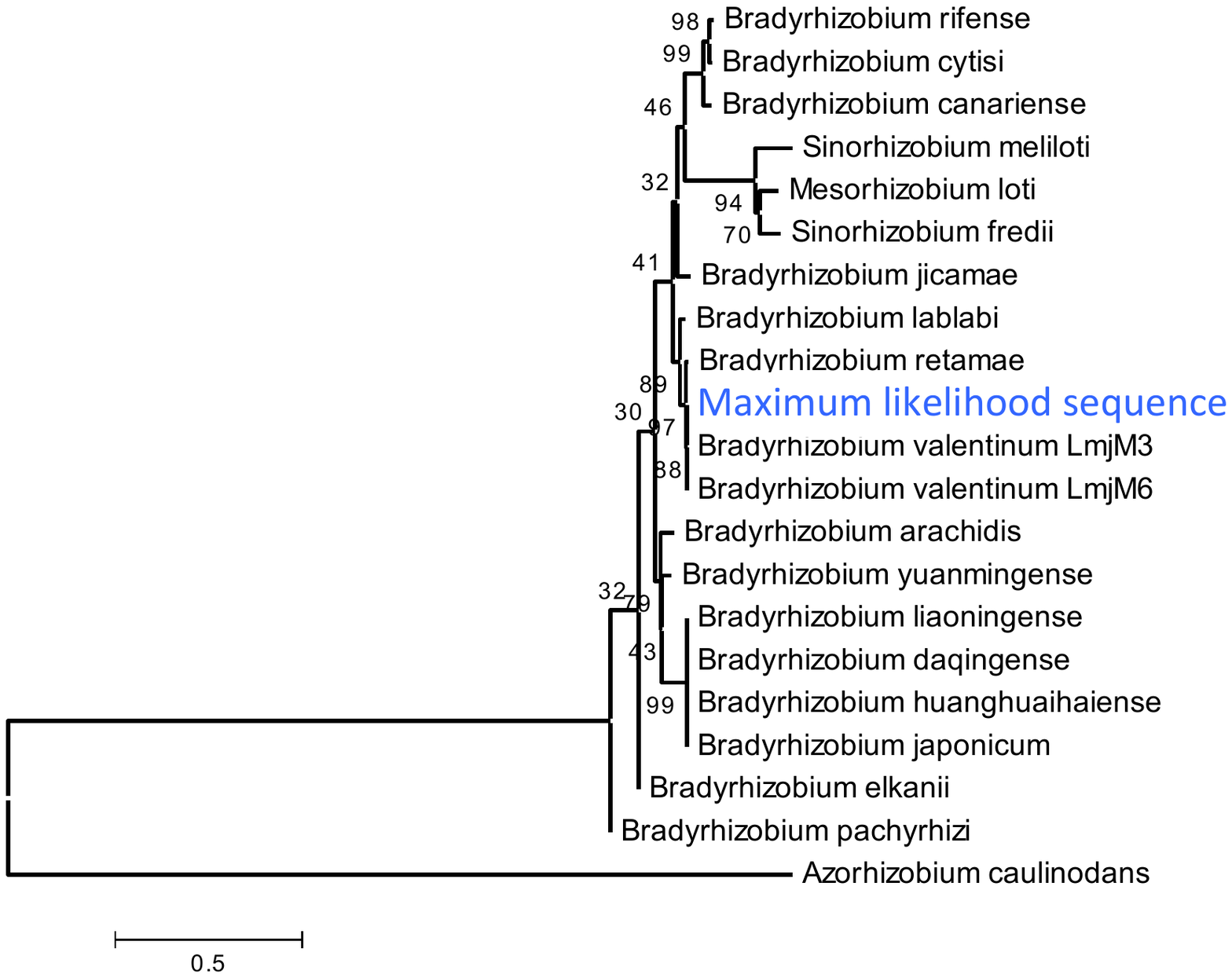}
    \vspace{-3ex}
    \caption{Neighbor-joining tree based on \textit{nifH} for 20 organisms along with the maximum likelihood sequence obtained from aligning the soil data to the gene sequence for \textit{nifH} of \textit{B. valentinum LmjM3}. Bootstrap values are shown next to the branches and the bar indicates 0.5 nucleotide substitutions per site. }
    \label{fig:TreeLengths}
\end{figure}

Figure \ref{fig:TreeLengths} depicts a tree using the same method as just described, but with evolutionary distances computed using the Maximum Composite Likelihood method \cite{tamura2004prospects}. Units are the number of base substitutions per site.

\section{Additional Figures}
\label{Appendix:AdditionalFigures}
We provide here a number of additional output figures from {\sf MetaPalette} to demonstrate that the ability to correctly infer the presence of organisms related to, but absent from, the training database is not dependent on the particular kingdom/phyla/etc.\ used. Unless otherwise noted, a total of 50K simulated reads from the novel organisms were spiked into the HMP mock even community. Figures are included for Bacteria, Archaea, Eukaryota, and viruses.

%\begin{figure}[hp!]
%    \centering
%    \includegraphics[scale=.15]{Figures/Spiking/Burkholderia_multivorans1.png}
%    \caption{HMP spiking results for the bacterial species \textit{Burkholderia multivorans} with 24 training strains and one novel testing strain.}
%    \label{}
%\end{figure}

%\begin{figure}[hp!]
%    \centering
%    \includegraphics[scale=.15]{Figures/Spiking/Lactobacillus_reuteri1.png}
%    \caption{HMP spiking results for the bacterial species \textit{Lactobacillus reuteri} with 17 training strains and one novel testing strain. }
%    \label{}
%\end{figure}

\begin{figure}[hp!]
    \centering
\includegraphics[scale=.15]{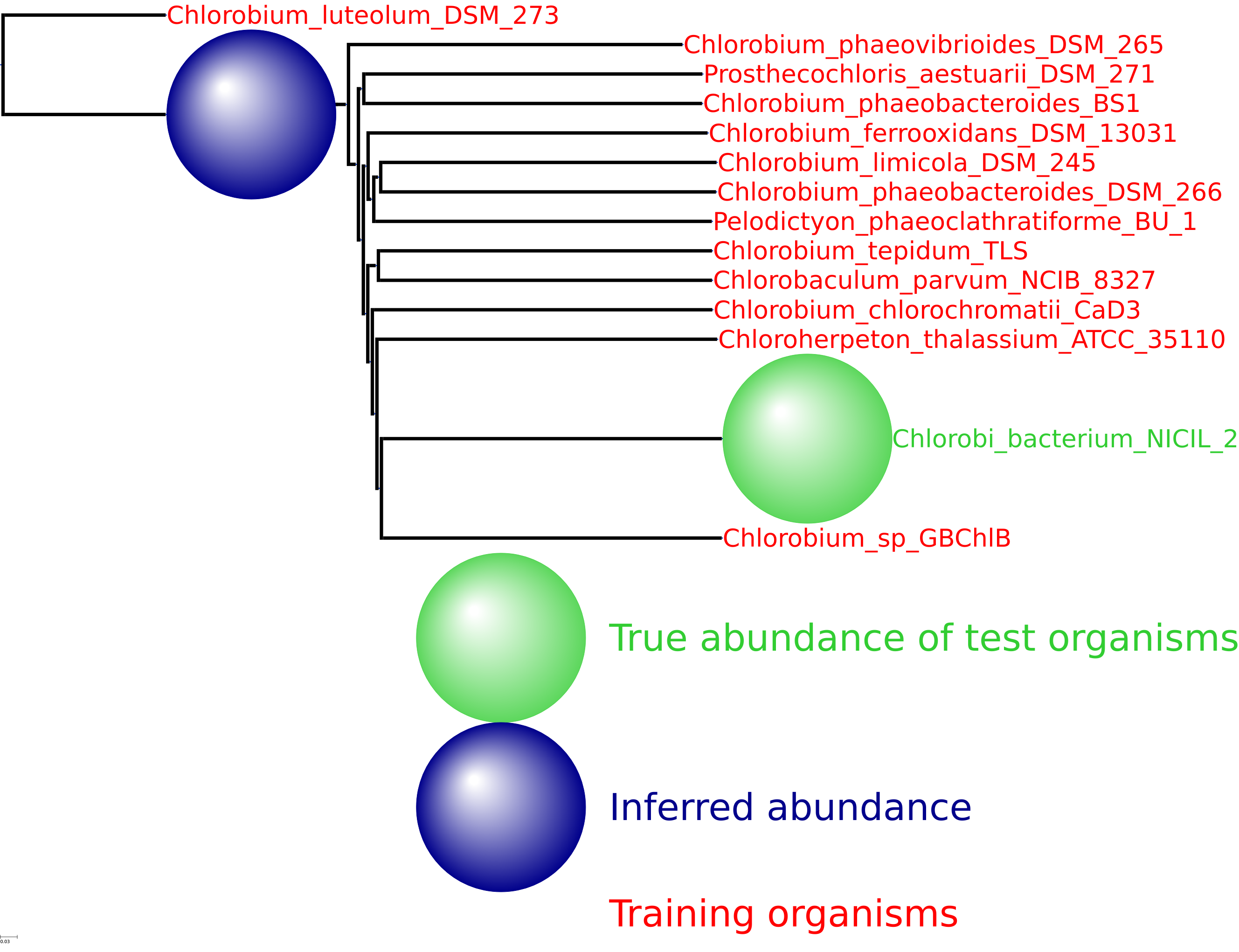}
    \caption{HMP spiking results for the bacterial phylum \textit{Chlorobi} with 13 training organisms and one novel testing organism.}
%    \label{}
\end{figure}

\begin{figure}[hp!]
    \centering
    \includegraphics[scale=.15]{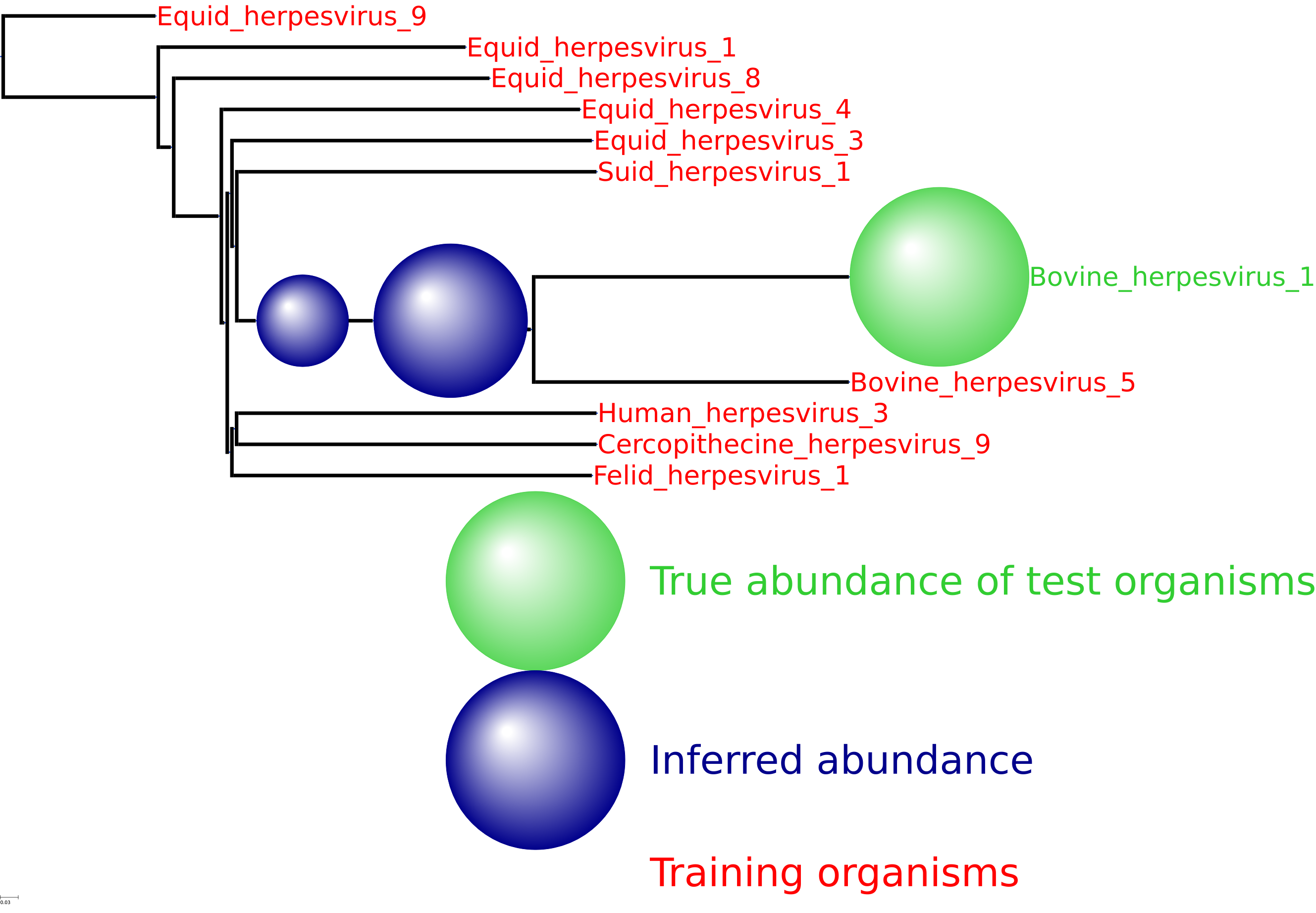}
    \caption{HMP spiking results for the viral genus \textit{Varicellovirus} with 10 training organism and one novel testing organism. Only 5K simulated reads were spiked into the HMP sample.}
%    \label{}
\end{figure}

\begin{figure}[hp!]
    \centering
    \includegraphics[scale=.15]{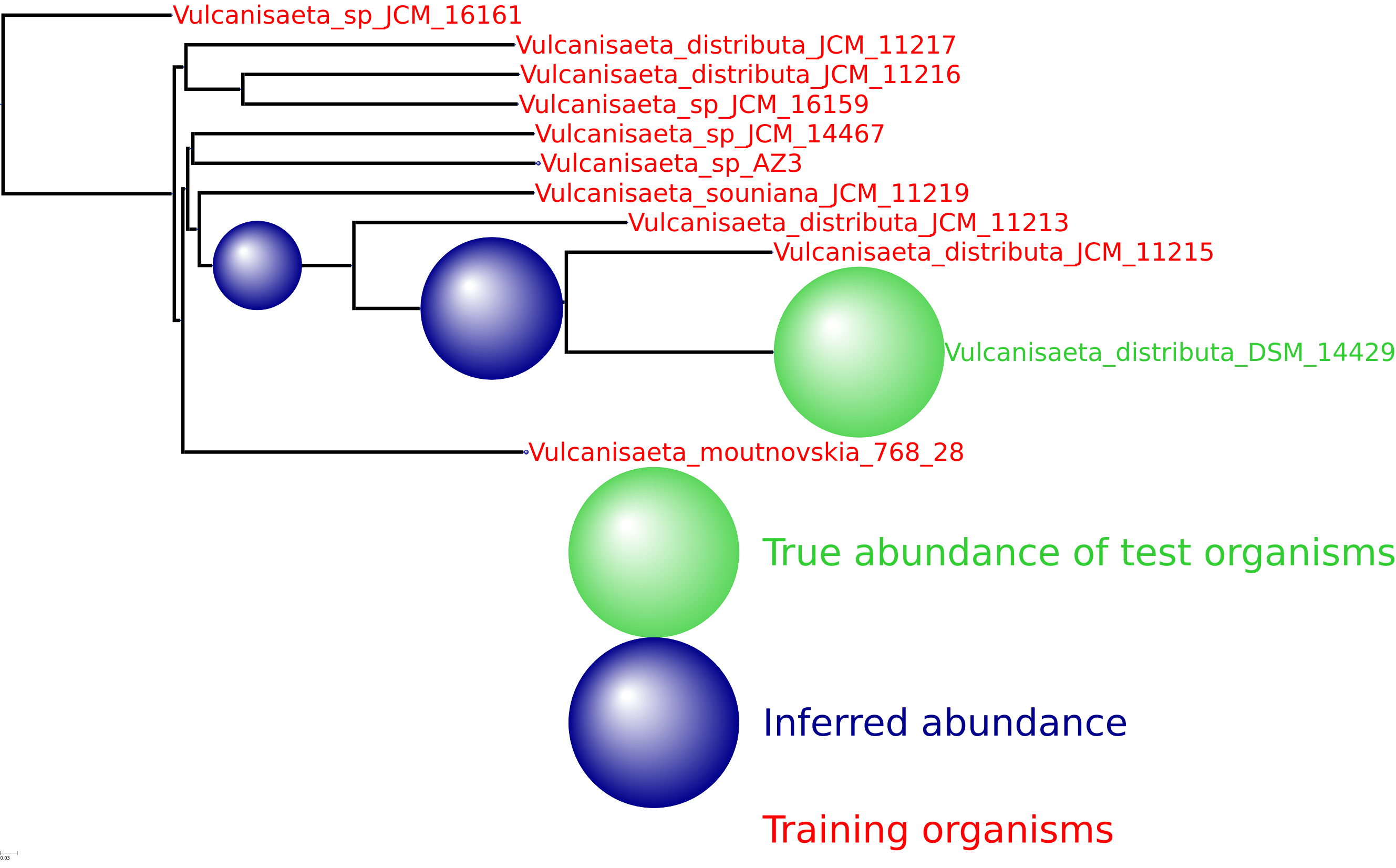}
    \caption{HMP spiking results for the archaeal genus \textit{Vulcanisaeta} with 10 training organism and one novel testing organism. Only 5K simulated reads were spiked into the HMP sample.}
%    \label{}
\end{figure}

\begin{figure}[hp!]
    \centering
    \includegraphics[scale=.10]{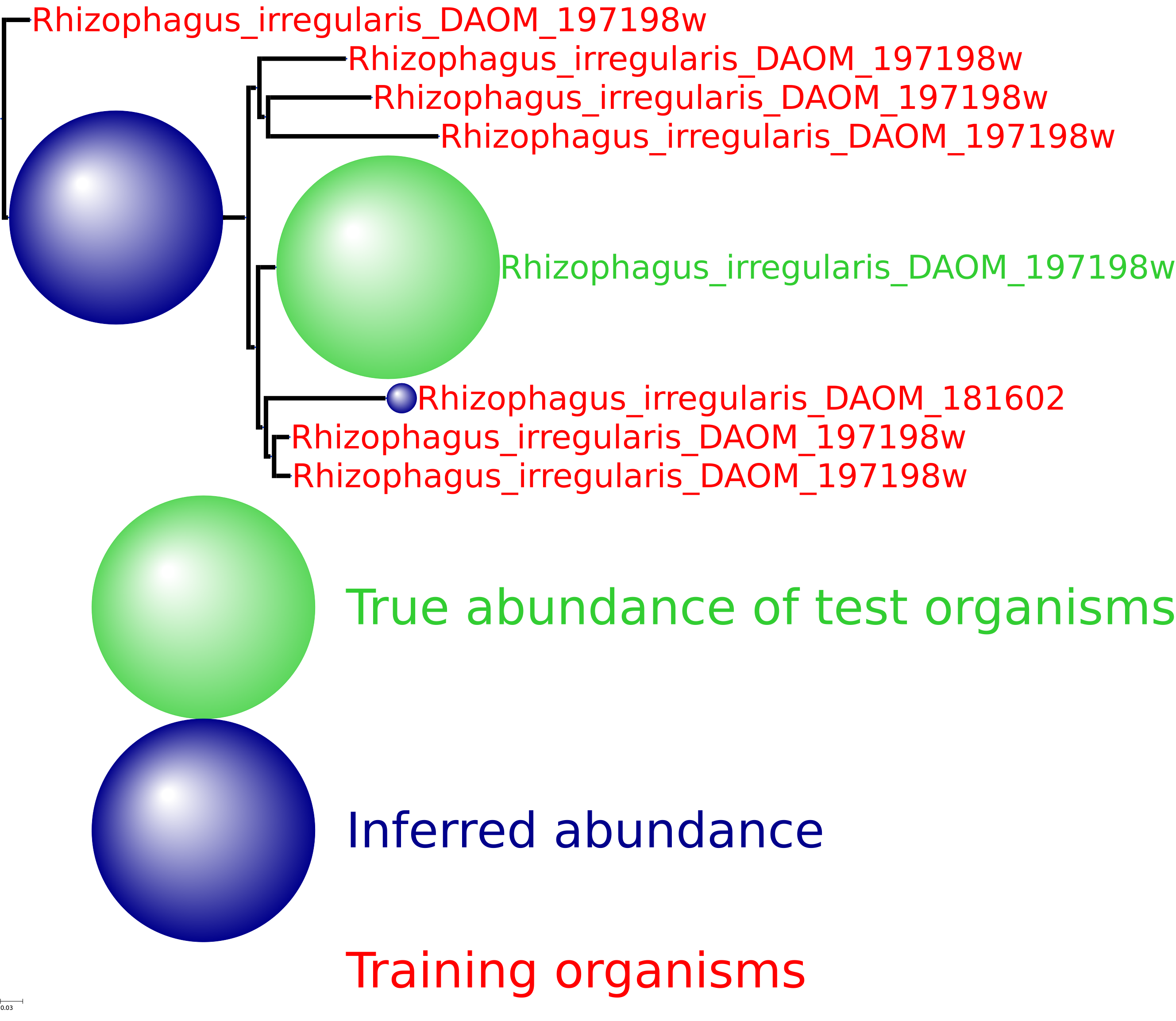}
    \caption{HMP spiking results for the eukaryota genus \textit{Rhizophagus} with 7 training organism and one novel testing organism.}
%    \label{}
\end{figure}

%\begin{figure}[hp!]
%    \centering
%    \includegraphics[scale=.15]{Figures/Spiking/Providencia_alcalifaciens1.png}
%    \caption{Here I spiked a real HMP data set (Mock even community, a mixture of 22 known organisms sequenced with an Illumina GA-II resulting in more than 6.5M reads ) with 50K reads of a missing organism and then ran MetaPalette. }
%    \label{}
%\end{figure}

%\begin{figure}[hp!]
%    \centering
%    \includegraphics[scale=.15]{Figures/Spiking/Providencia_alcalifaciens3.png}
%    \caption{Here I spiked a real HMP data set (Mock even community, a mixture of 22 known organisms sequenced with an Illumina GA-II resulting in more than 6.5M reads ) with 50K reads from two missing organisms and then ran MetaPalette. }
%    \label{}
%\end{figure}

%\begin{figure}[hp!]
%    \centering
%    \includegraphics[scale=.15]{Figures/Spiking/Lysinibacillus4.png}
%    \caption{HMP spiking results for the bacterial genus \textit{Lysinibacillus} with 30 training organism and three novel testing organism. }
%    \label{}
%\end{figure}

\clearpage
\bibliography{library}{}

\begin{thebibliography}{10}

\bibitem{ames2013scalable}
S.~K. Ames, D.~A. Hysom, S.~N. Gardner, G.~S. Lloyd, M.~B. Gokhale, and J.~E.
  Allen.
\newblock Scalable metagenomic taxonomy classification using a reference genome
  database.
\newblock {\em Bioinformatics}, 29(18):2253--2260, 2013.

\bibitem{Angly2012}
F.~E. Angly, D.~Willner, F.~Rohwer, P.~Hugenholtz, and G.~W. Tyson.
\newblock {Grinder: a versatile amplicon and shotgun sequence simulator.}
\newblock {\em Nucleic acids research}, 61(0):1--8, Mar. 2012.

\bibitem{baker2010enigmatic}
B.~J. Baker, L.~R. Comolli, G.~J. Dick, L.~J. Hauser, D.~Hyatt, B.~D. Dill,
  M.~L. Land, N.~C. VerBerkmoes, R.~L. Hettich, and J.~F. Banfield.
\newblock Enigmatic, ultrasmall, uncultivated archaea.
\newblock {\em Proceedings of the National Academy of Sciences},
  107(19):8806--8811, 2010.

\bibitem{blankenberg2010galaxy}
D.~Blankenberg, G.~V. Kuster, N.~Coraor, G.~Ananda, R.~Lazarus, M.~Mangan,
  A.~Nekrutenko, and J.~Taylor.
\newblock Galaxy: a web-based genome analysis tool for experimentalists.
\newblock {\em Current protocols in molecular biology}, pages 19--10, 2010.

\bibitem{caporaso2010qiime}
J.~G. Caporaso, J.~Kuczynski, J.~Stombaugh, K.~Bittinger, F.~D. Bushman, E.~K.
  Costello, N.~Fierer, A.~G. Pena, J.~K. Goodrich, J.~I. Gordon, et~al.
\newblock Qiime allows analysis of high-throughput community sequencing data.
\newblock {\em Nature methods}, 7(5):335--336, 2010.

\bibitem{chen1998}
S.~S. Chen, D.~L. Donoho, and M.~A. Saunders.
\newblock {Atomic Decomposition by Basis Pursuit}.
\newblock {\em SIAM Journal on Scientific Computing}, 20(1):33--61, Jan. 1998.

\bibitem{chikhi2014representation}
R.~Chikhi, A.~Limasset, S.~Jackman, J.~T. Simpson, and P.~Medvedev.
\newblock On the representation of de bruijn graphs.
\newblock In {\em Research in Computational Molecular Biology}, pages 35--55.
  Springer, 2014.

\bibitem{davenport2012genometa}
C.~F. Davenport, J.~Neugebauer, N.~Beckmann, B.~Friedrich, B.~Kameri,
  S.~Kokott, M.~Paetow, B.~Siekmann, M.~Wieding-Drewes, M.~Wienhöfer, S.~Wolf,
  B.~Tümmler, V.~Ahlers, and F.~Sprengel.
\newblock Genometa - a fast and accurate classifier for short metagenomic
  shotgun reads.
\newblock {\em PLoS ONE}, 7(8):e41224, 08 2012.

\bibitem{droge2015taxator}
J.~Dr{\"o}ge, I.~Gregor, and A.~McHardy.
\newblock Taxator-tk: precise taxonomic assignment of metagenomes by fast
  approximation of evolutionary neighborhoods.
\newblock {\em Bioinformatics}, 31(6):817--824, 2015.

\bibitem{duran2014bradyrhizobium}
D.~Dur{\'a}n, L.~Rey, A.~Navarro, A.~Busquets, J.~Imperial, and
  T.~Ruiz-Arg{\"u}eso.
\newblock Bradyrhizobium valentinum sp. nov., isolated from effective nodules
  of lupinus mariae-josephae, a lupine endemic of basic-lime soils in eastern
  spain.
\newblock {\em Systematic and applied microbiology}, 37(5):336--341, 2014.

\bibitem{felsenstein1985confidence}
J.~Felsenstein.
\newblock Confidence limits on phylogenies: an approach using the bootstrap.
\newblock {\em Evolution}, pages 783--791, 1985.

\bibitem{FoucartKoslicki2014}
S.~Foucart and D.~Koslicki.
\newblock Sparse recovery by means of nonnegative least squares.
\newblock {\em IEEE Signal Processing Letters}, 21(4):498--502, 2014.

\bibitem{freitas2015accurate}
T.~A.~K. Freitas, P.-E. Li, M.~B. Scholz, and P.~S. Chain.
\newblock Accurate read-based metagenome characterization using a hierarchical
  suite of unique signatures.
\newblock {\em Nucleic acids research}, page gkv180, 2015.

\bibitem{giardine2005galaxy}
B.~Giardine, C.~Riemer, R.~C. Hardison, R.~Burhans, L.~Elnitski, P.~Shah,
  Y.~Zhang, D.~Blankenberg, I.~Albert, J.~Taylor, et~al.
\newblock Galaxy: a platform for interactive large-scale genome analysis.
\newblock {\em Genome research}, 15(10):1451--1455, 2005.

\bibitem{goecks2010galaxy}
J.~Goecks, A.~Nekrutenko, J.~Taylor, et~al.
\newblock Galaxy: a comprehensive approach for supporting accessible,
  reproducible, and transparent computational research in the life sciences.
\newblock {\em Genome Biol}, 11(8):R86, 2010.

\bibitem{haft2012high}
D.~H. Haft and A.~Tovchigrechko.
\newblock High-speed microbial community profiling.
\newblock {\em Nature methods}, 9(8):793--794, 2012.

\bibitem{hellenthal2014genetic}
G.~Hellenthal, G.~B. Busby, G.~Band, J.~F. Wilson, C.~Capelli, D.~Falush, and
  S.~Myers.
\newblock A genetic atlas of human admixture history.
\newblock {\em Science}, 343(6172):747--751, 2014.

\bibitem{howe2014tackling}
A.~C. Howe, J.~K. Jansson, S.~A. Malfatti, S.~G. Tringe, J.~M. Tiedje, and
  C.~T. Brown.
\newblock Tackling soil diversity with the assembly of large, complex
  metagenomes.
\newblock {\em Proceedings of the National Academy of Sciences},
  111(13):4904--4909, 2014.

\bibitem{hunter2014ebi}
S.~Hunter, M.~Corbett, H.~Denise, M.~Fraser, A.~Gonzalez-Beltran, C.~Hunter,
  P.~Jones, R.~Leinonen, C.~McAnulla, E.~Maguire, et~al.
\newblock Ebi metagenomics—a new resource for the analysis and archiving of
  metagenomic data.
\newblock {\em Nucleic acids research}, 42(D1):D600--D606, 2014.

\bibitem{huson2011integrative}
D.~H. Huson, S.~Mitra, H.-J. Ruscheweyh, N.~Weber, and S.~C. Schuster.
\newblock Integrative analysis of environmental sequences using megan4.
\newblock {\em Genome research}, 21(9):1552--1560, 2011.

\bibitem{jaccard1901etude}
P.~Jaccard.
\newblock {\em Etude comparative de la distribution florale dans une portion
  des Alpes et du Jura}.
\newblock Impr. Corbaz, 1901.

\bibitem{kaper2004pathogenic}
J.~B. Kaper, J.~P. Nataro, and H.~L. Mobley.
\newblock Pathogenic escherichia coli.
\newblock {\em Nature Reviews Microbiology}, 2(2):123--140, 2004.

\bibitem{koslicki2013quikr}
D.~Koslicki, S.~Foucart, and G.~Rosen.
\newblock {Q}uikr: a method for rapid reconstruction of bacterial communities
  via compressive sensing.
\newblock {\em Bioinformatics}, page btt336, 2013.

\bibitem{koslicki2014wgsquikr}
D.~Koslicki, S.~Foucart, and G.~Rosen.
\newblock {WGSQ}uikr: fast whole-genome shotgun metagenomic classification.
\newblock {\em PloS one}, 9(3):91784, 2014.

\bibitem{lan2012using}
Y.~Lan, Q.~Wang, J.~R. Cole, and G.~L. Rosen.
\newblock Using the rdp classifier to predict taxonomic novelty and reduce the
  search space for finding novel organisms.
\newblock {\em PLoS one}, 7(3):e32491, 2012.

\bibitem{langmead2012fast}
B.~Langmead and S.~L. Salzberg.
\newblock Fast gapped-read alignment with bowtie 2.
\newblock {\em Nature methods}, 9(4):357--359, 2012.

\bibitem{lawson1974solving}
C.~L. Lawson and R.~J. Hanson.
\newblock {\em Solving least squares problems}, volume 161.
\newblock SIAM, 1974.

\bibitem{leslie2015fine}
S.~Leslie, B.~Winney, G.~Hellenthal, D.~Davison, A.~Boumertit, T.~Day,
  K.~Hutnik, E.~C. Royrvik, B.~Cunliffe, D.~J. Lawson, et~al.
\newblock The fine-scale genetic structure of the british population.
\newblock {\em Nature}, 519(7543):309--314, 2015.

\bibitem{lindgreen2015evaluation}
S.~Lindgreen, K.~L. Adair, and P.~Gardner.
\newblock An evaluation of the accuracy and speed of metagenome analysis tools.
\newblock {\em bioRxiv}, page 017830, 2015.

\bibitem{liu2010metaphyler}
B.~Liu, T.~Gibbons, M.~Ghodsi, and M.~Pop.
\newblock Metaphyler: Taxonomic profiling for metagenomic sequences.
\newblock In {\em Bioinformatics and Biomedicine (BIBM), 2010 IEEE
  International Conference on}, pages 95--100. IEEE, 2010.

\bibitem{marccais2011fast}
G.~Mar{\c{c}}ais and C.~Kingsford.
\newblock A fast, lock-free approach for efficient parallel counting of
  occurrences of k-mers.
\newblock {\em Bioinformatics}, 27(6):764--770, 2011.

\bibitem{marcy2007dissecting}
Y.~Marcy, C.~Ouverney, E.~M. Bik, T.~L{\"o}sekann, N.~Ivanova, H.~G. Martin,
  E.~Szeto, D.~Platt, P.~Hugenholtz, D.~A. Relman, et~al.
\newblock Dissecting biological “dark matter” with single-cell genetic
  analysis of rare and uncultivated tm7 microbes from the human mouth.
\newblock {\em Proceedings of the National Academy of Sciences},
  104(29):11889--11894, 2007.

\bibitem{merkel2014docker}
D.~Merkel.
\newblock Docker: lightweight linux containers for consistent development and
  deployment.
\newblock {\em Linux Journal}, 2014(239):2, 2014.

\bibitem{meyer2008metagenomics}
F.~Meyer, D.~Paarmann, M.~D'Souza, R.~Olson, E.~M. Glass, M.~Kubal, T.~Paczian,
  A.~Rodriguez, R.~Stevens, A.~Wilke, et~al.
\newblock The metagenomics rast server--a public resource for the automatic
  phylogenetic and functional analysis of metagenomes.
\newblock {\em BMC bioinformatics}, 9(1):386, 2008.

\bibitem{minot2015one}
S.~S. Minot, N.~Krumm, and N.~B. Greenfield.
\newblock One codex: A sensitive and accurate data platform for genomic
  microbial identification.
\newblock {\em bioRxiv}, page 027607, 2015.

\bibitem{nguyen2014tipp}
N.-p. Nguyen, S.~Mirarab, B.~Liu, M.~Pop, and T.~Warnow.
\newblock Tipp: taxonomic identification and phylogenetic profiling.
\newblock {\em Bioinformatics}, 30(24):3548--3555, 2014.

\bibitem{ounit2015clark}
R.~Ounit, S.~Wanamaker, T.~J. Close, and S.~Lonardi.
\newblock Clark: fast and accurate classification of metagenomic and genomic
  sequences using discriminative k-mers.
\newblock {\em BMC genomics}, 16(1):236, 2015.

\bibitem{philippe2011resolving}
H.~Philippe, H.~Brinkmann, D.~V. Lavrov, D.~T.~J. Littlewood, M.~Manuel,
  G.~W{\"o}rheide, and D.~Baurain.
\newblock Resolving difficult phylogenetic questions: why more sequences are
  not enough.
\newblock {\em PLoS Biol}, 9(3):e1000602, 2011.

\bibitem{puigbo2009search}
P.~Puigb{\`o}, Y.~I. Wolf, and E.~V. Koonin.
\newblock Search for a'tree of life'in the thicket of the phylogenetic forest.
\newblock {\em Journal of Biology}, 8(6):1, 2009.

\bibitem{rinke2013insights}
C.~Rinke, P.~Schwientek, A.~Sczyrba, N.~N. Ivanova, I.~J. Anderson, J.-F.
  Cheng, A.~Darling, S.~Malfatti, B.~K. Swan, E.~A. Gies, et~al.
\newblock Insights into the phylogeny and coding potential of microbial dark
  matter.
\newblock {\em Nature}, 499(7459):431--437, 2013.

\bibitem{roure2013impact}
B.~Roure, D.~Baurain, and H.~Philippe.
\newblock Impact of missing data on phylogenies inferred from empirical
  phylogenomic data sets.
\newblock {\em Molecular biology and evolution}, 30(1):197--214, 2013.

\bibitem{saitou1987neighbor}
N.~Saitou and M.~Nei.
\newblock The neighbor-joining method: a new method for reconstructing
  phylogenetic trees.
\newblock {\em Molecular biology and evolution}, 4(4):406--425, 1987.

\bibitem{salichos2013inferring}
L.~Salichos and A.~Rokas.
\newblock Inferring ancient divergences requires genes with strong phylogenetic
  signals.
\newblock {\em Nature}, 497(7449):327--331, 2013.

\bibitem{schaeffer2015pseudoalignment}
L.~Schaeffer, H.~Pimentel, N.~Bray, P.~Melsted, and L.~Pachter.
\newblock Pseudoalignment for metagenomic read assignment.
\newblock {\em arXiv preprint arXiv:1510.07371}, 2015.

\bibitem{sharpton2014introduction}
T.~J. Sharpton.
\newblock An introduction to the analysis of shotgun metagenomic data.
\newblock {\em Frontiers in plant science}, 5, 2014.

\bibitem{stoye1998rose}
J.~Stoye, D.~Evers, and F.~Meyer.
\newblock Rose: generating sequence families.
\newblock {\em Bioinformatics}, 14(2):157--163, 1998.

\bibitem{sunagawa2013metagenomic}
S.~Sunagawa, D.~R. Mende, G.~Zeller, F.~Izquierdo-Carrasco, S.~A. Berger, J.~R.
  Kultima, L.~P. Coelho, M.~Arumugam, J.~Tap, H.~B. Nielsen, et~al.
\newblock Metagenomic species profiling using universal phylogenetic marker
  genes.
\newblock {\em Nature methods}, 10(12):1196--1199, 2013.

\bibitem{tamura2004prospects}
K.~Tamura, M.~Nei, and S.~Kumar.
\newblock Prospects for inferring very large phylogenies by using the
  neighbor-joining method.
\newblock {\em Proceedings of the National Academy of Sciences of the United
  States of America}, 101(30):11030--11035, 2004.

\bibitem{tamura2013mega6}
K.~Tamura, G.~Stecher, D.~Peterson, A.~Filipski, and S.~Kumar.
\newblock Mega6: molecular evolutionary genetics analysis version 6.0.
\newblock {\em Molecular biology and evolution}, page mst197, 2013.

\bibitem{truong2015metaphlan2}
D.~T. Truong, E.~A. Franzosa, T.~L. Tickle, M.~Scholz, G.~Weingart, E.~Pasolli,
  A.~Tett, C.~Huttenhower, and N.~Segata.
\newblock Metaphlan2 for enhanced metagenomic taxonomic profiling.
\newblock {\em Nature methods}, 12(10):902--903, 2015.

\bibitem{wang2001limitations}
B.~Wang.
\newblock Limitations of compositional approach to identifying horizontally
  transferred genes.
\newblock {\em Journal of molecular evolution}, 53(3):244--250, 2001.

\bibitem{wood2014kraken}
D.~E. Wood and S.~L. Salzberg.
\newblock Kraken: ultrafast metagenomic sequence classification using exact
  alignments.
\newblock {\em Genome Biol}, 15(3):R46, 2014.

\end{thebibliography}
\bibliographystyle{abbrv}

\end{document}